\documentclass[%
 reprint,
nofootinbib,
 amsmath,amssymb,
 aps,
]{revtex4-2}

\usepackage{aas_macros}
\usepackage[normalem]{ulem}
\usepackage{color}
\usepackage{graphicx}% Include figure files
\usepackage{dcolumn}% Align table columns on decimal point
\usepackage{bm}% bold math
\usepackage{setspace}
\usepackage{etoolbox}
\usepackage{multirow}
\usepackage{subcaption}
\usepackage{hyperref}% add hypertext capabilities
\begin{document}

\preprint{APS/123-QED}

\title[Bayesian analysis for rotational curves with $\ell$-boson stars as a dark matter component]{Bayesian analysis for rotational curves with $\ell$-boson stars as a dark matter component}
\author{Atalia Navarro-Boullosa.}
\email{atalia.navarro.boullosa@gmail.com}
\affiliation{Departamento de F\'isica, DCI, Campus Le\'on, Universidad de Guanajuato, C.P. 37150, Le\'on, Guanajuato, M\'exico}
 \affiliation{Instituto de Ciencias F\'isicas, Universidad Nacional Aut\'onoma de M\'exico, Apdo. Postal 48-3, 62251 Cuernavaca, Morelos, M\'exico.}
\author{Argelia Bernal.}
\affiliation{Departamento de F\'isica, DCI, Campus Le\'on, Universidad de Guanajuato, C.P. 37150, Le\'on, Guanajuato, M\'exico} 
\author{J. Alberto Vazquez.}
\email{javazquez@icf.unam.mx}
\affiliation{Instituto de Ciencias F\'isicas, Universidad Nacional Aut\'onoma de M\'exico, Apdo. Postal 48-3, 62251 Cuernavaca, Morelos, M\'exico.}

\date{\today}% It is always \today, today,

%%===================================================================================================%%
\begin{abstract}
%%===================================================================================================%%

Using Low Brightness Surface Galaxies (LBSG) rotational curves we inferred the free parameters of $\ell$-boson stars as a dark matter component. The $\ell$-boson stars are numerical solutions to the non-relativistic limit of the Einstein-Klein-Gordon system, the Schrödinger-Poisson (SP) system. These solutions are parametrized by an angular momentum number $\ell = (N-1)/2$ and an excitation number $n$. We perform a bayesian analysis by modifying the SimpleMC code to perform the parameter inference, for the cases with $\ell = 0$, $\ell = 1$ and multi-states of $\ell$-boson stars. We used the Akaike information criterion (AIC), Bayesian information criterion and the Bayes factor to compare the excited state ($\ell$=1) and the multi-state case with the ground state ($\ell$=0) as the base model due to its simplicity. We found that the data in most galaxies in the sample  favours the multi-states case and that the scalar field mass tends to be slightly bigger than the ground state case.
\end{abstract}

%Use showkeys class option if keyword
\maketitle

%\tableofcontents

%%=------------------------  Section -------------------------------------------------------
\section{Introduction}\label{sec:Intro}
%%=------------------------ -------------------------------------------------------
Since the first high resolution observations of M31 by Vera Rubin and Kent Ford in 1970 \cite{1970RubinAndFord}, the galaxy rotation curves  have become an important observable for testing dark matter models. Most recently the authors in \cite{Pahwa_2018}, and references therein, have shown that Low Surface Brightness galaxies (LSBG) are suited candidates to test dark matter models due to their low visibility in the optical and HI photometry, therefore it is reliable to assume that rotational curve dynamics in those galaxies depends mainly on the dark matter component.
\\
Throughout the years there has been a plethora of proposals about the nature of this mysterious component, being the scalar field dark matter (SFDM) a candidate that have been proved to be viable and that display several advantages over the standard candidate $\Lambda$ Cold Dark Matter ($\Lambda$CDM). 
For instance, one point in favor of the SFDM is that it does not suffer of the so-called ``cusp/core" problem, -- which states that the predicted density profiles, from simulations with $\Lambda$CDM, increase steeply and hence producing a ``cuspy" dark matter distribution in small radii, while observations of dark matter density profiles for most of the dwarf galaxies indicate otherwise: they display flat central ``cores" \cite{2010Core/Cusp,2001Core/Cusp} --. 
Another possible discrepancy between observations and simulations with $\Lambda$CDM is the missing satellite problem (MSP). Simulations indicate the presence of more dwarf galaxies orbiting around a galaxy with the characteristics of the Milky Way, than those we are able to observe around us. In the past years some discussions have taken place about the MSP and how the high resolution observations and $\Lambda$CDM simulations could give evidence of a plausible solution, for more details consult \cite{2018NOMSP}.
One of the small scale discrepancies coming from the $\Lambda$CDM simulations is the ``too big to fail" problem.  This states that the galaxy satellites predicted by the model are too massive that it is impossible that they doesn't have visible stars, meaning that, the observed satellites of the Milky Way are not massive enough to be consistent with predictions from $\Lambda$CDM \cite{2017Breview}.
\\
On the other hand, the SFDM model, -- which in the simplest case, considers that dark matter can be modeled by a classical, spin-zero, massive, minimally coupled to gravity, complex scalar field -- reproduces, at cosmological scales, the $\Lambda$CDM, background and linear density perturbations dynamics. An important difference is that the bosonic mass, which is the only free parameter of the model, fix a natural cut-off in the mass power spectrum at small scales \cite{Matos_2001}, preventing the overproduction of small structures. The mass of the scalar field needs to be larger than $10^{-23}$ eV in order to be compatible with power spectrum measurements \cite{2015PhRvD..91j3512H,2016JCAP...07..048U}.\\
At galactic scales, such a small mass implies cored galactic halos instead cuspy as $\Lambda$CDM produces. Structure formation simulations of SFDM \cite{2014Schive} provide density profiles for dark matter halos with a soliton core and an outer Navarro-Frenk-White profile \cite{1996NFW}. This density profile has been used to constrain dark matter features by using different systems such as the dwarf spheroidal galaxies (dSphs) \cite{2017Alma,pozo2021detection}. The analysis has shown that a single state of the SFDM tightly constrained the scalar field mass, $m$,  with Lyman-$\alpha$ observations  of about ($\log_{10}$($m$/eV) $\sim$ $[$-23,-24$]$) \cite{PhysRevLett.119.031302_Lymanalpha_constrain,2019Lyman_Alpha_constrain}. 
Also, an analysis of different density profiles, by using non-parametric reconstruction of rotational curves, showed that 44\% of the sample galaxies preferred the SFDM model \cite{2019MNRAS.488.5127F}.
\\
As an effort to extend this scalar field dark matter approach, it has been suggested fields with self-interactions  \cite{PhysRevLett.57.2485,1998PhRvD..58j4004B}, axionic-like potential has been widely studied from a pure gravitational approach in \cite{2017PhRvD..96f1301C,2021JCAP...01..051L}, see also references therein. Similarly, in \cite{2012JCAP...11..024P} they used the self-interacting term in the lagrangian and LBSG to constrain the Bose-Einstein condensate (BEC) dark matter within the range ($10^{-6}$, $10^{-4}$) eV, while in \cite{2013PhRvD..88l3517B} dark matter is composed of ordinary QCD axions,and the fact that QCD axions form a BEC is a consequence of their properties studied.\\
Phenomenologically, dark matter described by a classical ultralight scalar field has shown to be a good candidate of dark matter, however the approach of this matter within the quantum field theory considering the quantum nature of the possible fundamental boson associated with it, has been less explored.  A hint could come from Boson Stars (BS).
BS where introduced by Kaup \cite{PhysRev.172.1331} as regular localized solutions of the Einstein Klein-Gordon equations for a massive minimally coupled classical complex scalar field. Soon after, Ruffini and Bonazzola presented them as static spherically symmetric self-gravitating configurations of quantum spin-zero particles in quantum field theory using the semiclassical gravity approximation \cite{PhysRev.187.1767}. For a general overview about boson stars we recommend \cite{2012LRR....15....6L,2021IJMPD..3030006V}. Ruffini and Bonazzola's solutions coincide with the classical configurations  if all particles populate a specific energy state however, the semiclassical approximation unveil more sophisticated stars. In particular, even within the spherical symmetry and staticity, configurations in which different energy states are populated simultaneously, naturally emerge. Such kind of configurations, in Kaup's approach,  require introducing multiple independent classical fields, one for each occupied state, while in the semiclassical approximation different states are excitations of a single quantum field \cite{2023PhRvD.107d5017A}. 
\\
 BS with all the particles in the ground state are viable as astrophysical objects, they are stable solutions \cite{PhysRevD.42.384}, and furthermore, they have been shown to be atractor solutions under fairly general initial conditions \cite{2006PhRvD..74f3504B,2023arXiv230403419A}. In the SFDM scenario they would be the kind of structures formed in isolation boundary conditions \cite{2023arXiv230403419A,2019arXiv190708193G} and have been used to model SFDM galactic halos to put constrains on the bosonic mass using different observables such as galactic rotation curves (RC) \cite{2015JPhCS.640a2056B} and stellar cluster dynamics \cite{2012JCAP...02..011L}. For halos modeled by BS in the ground state, it is known that RC of different galaxies require different masses of the boson to be fitted, see for instance results for LSBG \cite{2015JPhCS.640a2056B}. It is possible to reproduce RC of different galaxies with the same bosonic mass using configurations in excited states (all the bosons not in the ground state), however those configurations are unstable \cite{2003PhRvD..68b4023G}. 
\\
Configurations with bosons populating different states provide a diversity of mass density profiles that could account for the dark matter distribution in galaxies with diverse characteristics. In such configurations each occupied state is labeled by quantum numbers {\it n}, $\ell$ and {\it m}, ground states mentioned in the above paragraph have  {\it n} $=0$, $\ell=m=0$. In \cite{2007GReGr..39.1279M}, configurations with $n\neq 0$, $\ell=m=0$, for which {\it n} takes different values, the so called "multi-state" configurations, were proposed to obtain flat rotation curves within the SFDM. An important condition for the physical viability of those configurations is its stability, in \cite{2010PhRvD..82l3535U,PhysRevD.81.044031} it was proven that there exist stable and virialized multi-state configurations. \\
In this work, we explore rotation curves obtained from spherically symmetric self gravitating configurations with particles in states for which the angular momentum is  non-zero. The simplest case of those configurations are the so called $\ell$-boson stars, in them, all particles have fixed radial and total angular momentum numbers $n$ and $\ell$, with $\ell \neq 0$, but are homogeneously distributed with respect to their magnetic number $m$. In \cite{Alcubierre_2018} they were introduced in the classical regime as configurations composed by $N=2\ell+1$ classical independent complex scalar fields with the same mass. Each of these fields have a non trivial harmonic with an angular momentum number $m=-\ell$, $-\ell+1$, \ldots, $\ell$, but all of them have the same temporal and radial dependence. It is interesting that this special combination gives a static configuration with zero total angular momentum even when independently the fields have angular momentum and are time dependent. At this point, in the classical approximation, the existence of different fields combined with such special characteristics could be seen as somehow artificial. 
This is different within the semiclassical approximation, where the same $\ell$-boson star is an allowed state of  a single, massive, real, free quantum scalar field. The state describes the excitation of $N$ excitation modes of the quantum field. The corresponding Einstein-Klein-Gordon system of equations, considering the expectation value of the stress energy-momentum tensor operator takes the same form as its counterpart in a classical theory with $N$ independent complex fields \cite{2023PhRvD.107d5017A},  (see also reference  \cite{Guzm_n_2020} in which the Newtonian limit of those configurations is studied). In the same sense the "multi-$\ell$-boson stars", used to model dark matter halos in this work are allowed states of the quantum field. We consider configurations with $n = \ell +1$, $m=-\ell$, $-\ell+1$, \ldots, $\ell$, for which $\ell$ takes different values and we take a phenomenological point of view, observations will telling us which configurations are preferred. To find a more fundamental explanation about which modes of the quantum field should be excited is an open, interesting and necessary research field.\\
The stability of self-gravitating scalar field configurations have been widely studied as it is an important characteristic to determine their viability as astrophysical objects. In \cite{2021CQGra..38q4001A,2019CQGra..36u5013A} and \cite{PhysRevD.107.084001,https://doi.org/10.1002/asna.202113941} it was shown that there exist stable $\ell$-boson stars with $n=\ell +1$, under radial perturbations, in the relativistic and Newtonian limit respectively. On the other hand, $\ell$-boson stars are unstable under 3D perturbations, however it is possible to stabilize them by adding a sufficiently large fundamental, $n=1$, $\ell = 0$ boson star \cite{2021PhRvL.126x1105S}. \\
Multi-$\ell$-boson stars are allowed states of the quantum field that contain information of the angular momentum preserving the spherical symmetry, its 3D stability is expected if $n=\ell + 1$ and if the contribution of the $\ell=0$ state is large enough, furthermore their density profile differ significantly from their relatives with $\ell=0$, in particular for $\ell>1$ the central density of $\ell$-boson stars is zero while for stars with $\ell=0$ the central density is a global maximum. These characteristics make multi-$\ell$-boson stars suitable to be tested as part of the galactic halo, in this work we only considered configurations with $n=\ell + 1$ and we found that best fits of RC are possible with multi-$\ell$-boson stars in which the $\ell = 0$ component is important, then those multi-$\ell$-boson stars modeling dark matter galactic halos should be stable although a definite proof could be done in a future work.\\
This paper is organized as follows, in section \ref{Sec:Model} we present the bases of the SFDM model in the non-relativistic limit, their characteristics and the three specific cases that we analyzed: the ground state or boson star ($\ell=0$), a boson star with an excited state and multi-states.
In \ref{Sec:Data} we describe the galaxy data set used with the methods described in section \ref{Sec:Analysis}. Our results, presented in section \ref{Sec:Results}, showed that most of the galaxies have a better fit to the data by using  multi-states.
%, 
 Finally, in section \ref{Sec:Conclusions_and_discussion} we discuss our results and perspectives. 

%%=------------------------  Section ----------------------------------------
\section{Model} \label{Sec:Model}
%%=----------------------------------------------------------------

We take an spin 0 scalar field without self-interaction and 
use the non-relativistic limit of the Einstein-Klein Gordon system of equations as mentioned in \cite{Guzm_n_2020}. For the non-relativistic $\ell$-boson stars configurations the Schödinger-Poisson (SP) system is
\begin{subequations}
\begin{equation}
  \nabla_{r\ell}^{2}V_{0 0}= 4\pi Gm_{a}^{2}\sum_{n, \ell}\left(2 \ell+1\right) r^{2 \ell} \psi_{n \ell 0}^{2},
\end{equation}
and
\begin{equation}
    \frac{\hbar^{2}}{2m_{a}}\nabla_{r \ell}^{2} \psi_{n \ell 0}=\left(m_{a}V_{00}-\gamma_{n \ell 0}\right) \psi_{n \ell 0}.
    \label{eq:10b}
\end{equation}
\end{subequations}
Where $G$ is the gravitational constant, $m_{a}$ is the mass of the scalar field, $\hbar$ is the reduced Planck constant, $\gamma_{n\ell 0}$ is the frequency that will be determined by solving the eigenvalue problem, described in subsection \ref{subsec:100,210,320}, $V_{00}$ is the gravitational potential with spherical symmetry, the subscript indicates that the only term different to zero in the multipolar expansion is the monopolar one, which can be consulted in \cite{Guzm_n_2020}; and 
\begin{equation}
    \nabla_{r \ell}^{2} = \partial^{2}_{r} + [2(\ell +1)/r] \partial_{r}.
\end{equation}
By using the following expressions 
\begin{equation}
\begin{aligned}
    \psi_{n\ell 0} = \bar{\psi}_{n\ell 0}\frac{\epsilon^{2}c^{2}}{\hbar\sqrt{4\pi G}}, \qquad
    r = \frac{\bar{r}}{\epsilon}\frac{\hbar}{m_{a}c}, \qquad
     V_{00} = \bar{V}_{00}\epsilon^{2}c^{2},
\end{aligned}
\label{eq:cambios1}
\end{equation}
we obtain a set of dimensionless equations, where the bar indicates the numerical solution, $c$ is the speed of light and $\epsilon$ is a dimensionless quantity related with the speed of light and the rescaling amplitude for our rotational curve, being one of the free parameters of our model along with the mass of the scalar field. For simplicity, we omit the bar notation, giving us the following system of equations  
\begin{subequations}
\begin{equation}
     \frac{d^{2}\psi_{n\ell 0}}{dr^{2}} = - \frac{2(\ell+1)}{r} \frac{d\psi_{n\ell 0}}{dr} + 2(V_{00}-\gamma_{n\ell 0})\psi_{n\ell 0},
 \label{eq:dif_Schrodinger}
\end{equation}
and
\begin{equation}
    \frac{d^{2}V_{00}}{dr^{2}} = -\frac{2}{r}\frac{dV_{00}}{dr} + \sum_{n,\ell}(2\ell +1)r^{2\ell}\psi_{n\ell 0}^{2}.
    \label{eq:dif_Poisson}
\end{equation}
\end{subequations}
To obtain the numerical solutions to the equation system, we add the following expression for the number of particles in each state \cite{2010PhRvD..82l3535U}
\begin{equation}
    \frac{dN_{\ell}}{dr} = \psi_{n\ell 0}^{2}r^{2+2\ell},
    \label{eq:dNumber_of_particles}
\end{equation}
and implemented the shooting method with a fourth order Runge Kutta. 
\\
Because the system of equations only depends of the radial coordinate, we can use the expression for the circular speed ($V_{c}$) with spherical symmetry to obtain the rotational curve for the non-relativistic $\ell$-boson stars
\begin{equation}
    V_{c}^{2} (r) = \frac{GM(r)}{r},
\end{equation}
where we used the change of variables for the dimensionless solutions (\ref{eq:cambios1}) and performed the numerical integral from $0$ to $R$ to obtain the mass $M(r)$ function
\begin{equation}
    M(r) = 4\pi \int_{0}^{R}\rho(r)r^{2}dr.
   \label{eq:M(r)general}
\end{equation}
In the above expression the density profile is given by 
\begin{equation}
\rho(r) = \frac{m_{a}^{2}\epsilon^{4}}{4\pi(1.95\times 10^{-69})}\sum_{n,\ell} (2\ell +1)r^{2\ell}\psi_{n\ell 0}^{2}\frac{M_{\odot}}{\text{kpc}^{3}},
\label{eq:rho_sum}
\end{equation}

\noindent 
where $M_{\odot}$ means solar masses. Therefore the circular speed ($V_{c}$) in terms of the new variables (\ref{eq:cambios1}) is
\begin{equation}
    V_{c}^{2} = 8.95\times 10^{10} \frac{\epsilon^{2}}{R}\int_{0}^{R} r^{2}\sum_{n,\ell} (2\ell +1)r^{2\ell}\psi_{n\ell 0}^{2} dr\left(\frac{\rm km}{\rm s}\right)^{2}.
    \label{eq:Vc2_general}
\end{equation}
Notice that the above equation only depends on $\epsilon$ and the radial distance has a dependency on $\epsilon$ and $m_{a}$ (see (\ref{eq:cambios1})).
\\
In particular, we choose three cases of states with zero nodes to assure stability, following the selection rule $n-1-\ell = 0$ \cite{Guzm_n_2020}: the ground state with $\ell = 0$ (\ref{subsec:100}),  an excited state with $\ell =1$ (\ref{subsec:210}) and the multi-state with $\ell = 0,1,2$ (\ref{subsec:100,210,320}). In the next subsections we present the boundary conditions to solve the system of equations of each case and their characteristics (\ref{subsec:characteristics}).

%%---------------------------------------------------------------------------------
\subsection{Ground state: $\psi_{100}$} \label{subsec:100}
%%---------------------------------------------------------------------------------

We solved the dimensionless equation system given by (\ref{eq:dif_Schrodinger}) and (\ref{eq:dif_Poisson}), with $\ell = 0$ and $n=1$, better known as a simple boson star, this solutions can be seen in the FIG \ref{fig:family_100}; this case has been well studied in \cite{2007JCAP...06..025B,2003PhRvD..68b4023G}.
The boundary conditions to be determined by the shooting method are: $\gamma_{100}(r=0)$ and $V_{00}(r=0)$. 
It is important to mention that for the independent cases, i.e. the ground state \ref{subsec:100} and the excited state \ref{subsec:210}, $\psi_{100}(r=0)$ is fixed to one, this is due to the rescaling properties discussed in subsection \ref{subsec:characteristics} and the fact that the solutions form a family allowed us to use the parameter $\epsilon$ as the rescaling factor in the initial amplitude ($\psi_{100}(r=0)$). Therefore, the free parameters to be estimated for these cases are $\epsilon$ and $m_{a}$.
%

%%---------------------------------------------------------------------------------
\subsection{Excited state: $\psi_{210}$}\label{subsec:210}
%%---------------------------------------------------------------------------------

Considering the case where the dark matter halo is only form by an excited state, which is related to the quadrupole symmetry, we solved the equation system given by (\ref{eq:dif_Schrodinger}) and (\ref{eq:dif_Poisson}), with $\ell = 1$ and $n=2$ different to zero. The boundary conditions are the same as the ones for the ground state \ref{subsec:100}, where $\gamma_{210}(r=0)$ and $V_{00}(r=0)$  will be determined by solving the eigenvalue problem.

%%---------------------------------------------------------------------------------
\subsection{Multi-states: $\psi_{100}$, $\psi_{210}$, $\psi_{320}$}\label{subsec:100,210,320}
%%---------------------------------------------------------------------------------

As it was mentioned before, we choose the states with zero nodes to assure stability and followed the selection rule $n-1-\ell = 0$ \cite{Guzm_n_2020}. For the multi-state case we took $\ell = 0,1,2$. Therefore, the system of equations to be solved numerically is
\begin{subequations}
\begin{equation}
    \frac{d^{2} \psi_{100}}{d r^{2}}=2\left(V_{00}-\gamma_{100}\right) \psi_{100}-\frac{2}{r} \frac{d \psi_{100}}{d r},
    \label{eq:psi_100_2}
\end{equation}
\begin{equation}
    \frac{d^{2} \psi_{210}}{d r^{2}}=2\left(V_{00}-\gamma_{210}\right) \psi_{210}-\frac{4}{r} \frac{d \psi_{210}}{d r},
    \label{eq:psi_110_2}
\end{equation}
\begin{equation}
    \frac{d^{2} \psi_{320}}{d r^{2}}=2\left(V_{00}-\gamma_{320}\right) \psi_{320}-\frac{6}{r} \frac{d \psi_{320}}{d r},
    \label{eq:psi_320}
\end{equation}
\begin{equation}
   \frac{d^{2} V_{00}}{d r^{2}}=-\frac{2}{r} \frac{d V_{00}}{d r}+\psi_{100}^{2}+3 r^{2} \psi_{210}^{2}+5 r^{4} \psi_{320}^{2}.
   \label{eq:V00_012}
\end{equation}
\end{subequations}

Boundary conditions must guarantee that the solutions are regular and asymptotically flat. This implies that \ref{eq:V00_012} becomes an eigenvalue problem for $\gamma_{100}$, $\gamma_{210}$ and $\gamma_{320}$. 
Regularity at the origin implies $\psi_{100}(r=0)=C_1$, $\psi_{210}(r=0)=C_2$, $\psi_{320}(r=0)=C_3$ and $\psi_{i00}'(r=0)=0$, for the potential $V_{00}'(r=0)=0$. Asymptotically flatness implies $\psi_{i00}(r\to \infty)\to 0$ and we impose in this boundary $V_{00}/r+V_{00}'=0$.\\
To find the solutions we solved the eigenvalue problem to determine the frequencies $\gamma_{100}$, $\gamma_{210}$ and $\gamma_{320}$ by using a shooting method. 
We took the Taylor expansions around zero to obtain the boundary conditions, finding that $\psi_{n\ell 0}''(0) = \frac{2V_{00}(0)\psi_{n\ell 0}(0)}{3+2\ell}$ and to avoid divergences we cancel out the terms $1/r$   for the integration in $r=0$. 
The first integration  starts from $r=0$ to a close boundary and after finding a good frequency guess, the new frequency value is taken to the next step in the integration, until the frequencies converge to a desired precision.
Therefore, $\psi_{100}$(0), $\psi_{210}$(0) and $\psi_{320}$(0) became free parameters 
along with $\epsilon$ and $m_{a}$. It is important to mention that adding the states central amplitudes to the free parameters implies the resolution of the eigenvalue problem for every step in the sampling algorithm, this makes the algorithm more computationally expensive.
The expression for the number of particles in each state (\ref{eq:dNumber_of_particles}) is part of the system of equations, meaning that the solutions for $N_{\ell =0}$, $N_{\ell =1}$ and $N_{\ell =2}$ are being found.

%%---------------------------------------------------------------------------------
\subsection{The characteristics of the solutions}\label{subsec:characteristics}
%%---------------------------------------------------------------------------------

\begin{figure}
    \centering
    \includegraphics[width=3in]{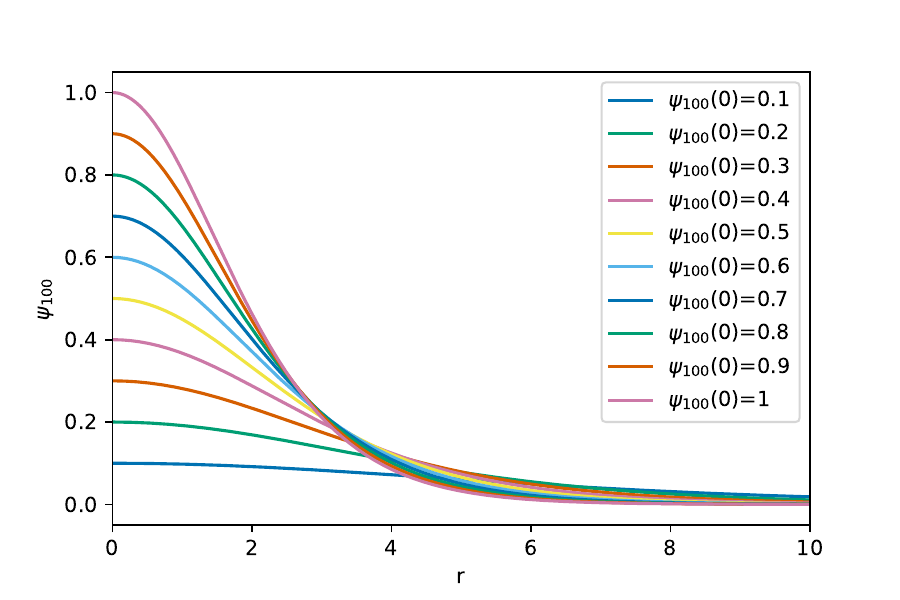}
    \caption{Numerical solutions for the system of equations described in section \ref{subsec:100} for the ground state. Each solution was obtained using the expression (\ref{eq:lambda}) with $\psi_{100} = 1$ as a source.}
    \label{fig:family_100}
\end{figure}

Solving numerically the SP equations we could observe interesting characteristics. Taking the independent cases ($\psi_{100}$ and $\psi_{210}$) we observe that they form a family, meaning that, using the solution for a given central amplitude $\psi_{n\ell 0}(0)$ we can use the expression
\begin{equation}
    \lambda = \left(\frac{1}{\psi_{n\ell 0}(0)}\right)^{(1/(\ell+2))},
    \label{eq:lambda}
\end{equation}
to obtain the solution with a different central amplitude $\psi_{n\ell 0}(0)$ without solving the system of equations once again, see FIG \ref{fig:family_100}. This is due to rescaling properties in the equations system that can be found using $(r,\psi_{n\ell 0 , V_{00}})\to (r\lambda,\psi_{n\ell 0}/\lambda^{\ell +2}, V_{00}/\lambda^{2})$. 
Therefore, for the independent solutions the parameter $\lambda$ is analogous to the parameter $\epsilon$, this equivalence is broken for the multi-state solutions, this can be seen in the equation (\ref{eq:rho_sum}), where the parameter $\epsilon$ is multiplying the sum over $n$ and $\ell$, therefore, if the parameter $\epsilon$ changes it will change all the states over the sum equally.
\\

\begin{figure}
\centering
\includegraphics[width=3in]{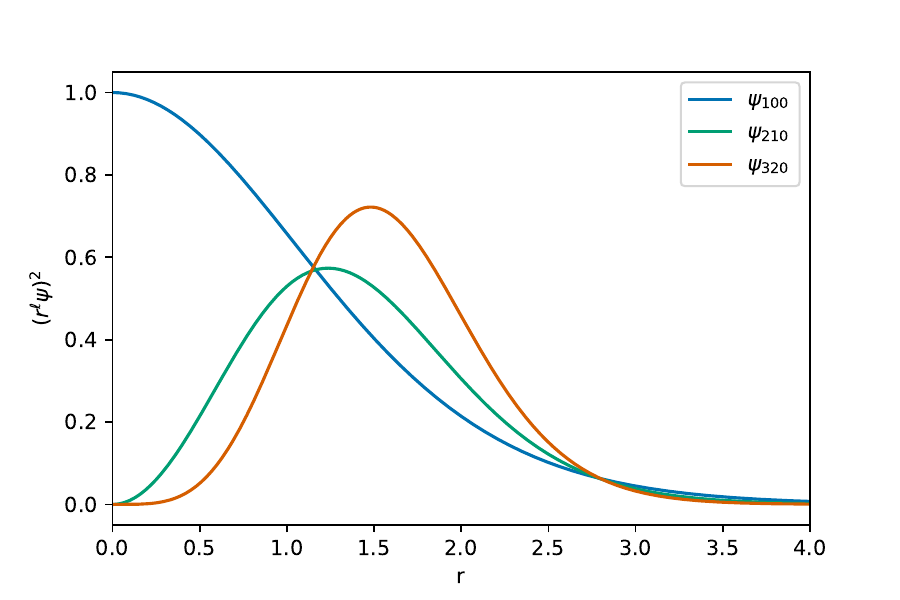}
\caption{Numerical density profiles of each state, solving the system of equations described in section \ref{subsec:100} for the ground state ($\psi_{100}$), described in section \ref{subsec:210} for the excited state ($\psi_{210}$) and its analogous for $\psi_{320}$. The blue line corresponds to the state $\psi_{100}$, the green line to the state $\psi_{210}$ and the orange line to the $\psi_{320}$ state.}\label{fig:density_profiles01_ind}
\end{figure}

\begin{figure}
\centering
\includegraphics[width=3in]{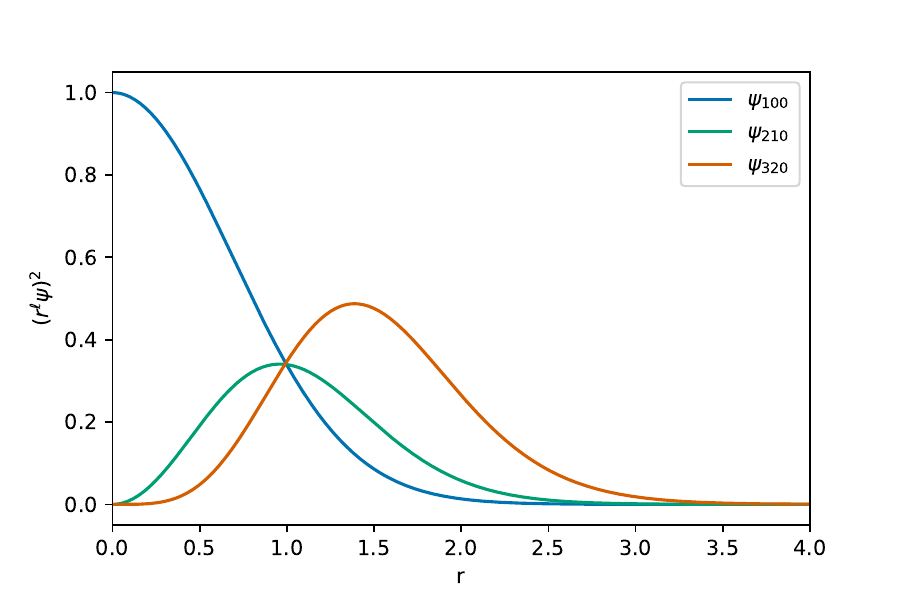}
\caption{Numerical density profiles of each state, solving the system of equations (\ref{eq:psi_100_2} - \ref{eq:V00_012}). The blue line corresponds to the state $\psi_{100}$, the green line to the state $\psi_{210}$ and the orange line corresponds to the state $\psi_{320}$.}\label{fig:density_profiles012}
\end{figure}

In FIG \ref{fig:density_profiles01_ind} and FIG \ref{fig:density_profiles012}, we show the numeric density profiles for the system of equations described in sections \ref{subsec:100}, \ref{subsec:210} and \ref{subsec:100,210,320}, respectively. 
Comparing these figures we can observe that in the FIG \ref{fig:density_profiles01_ind} each numerical density has a bigger amplitude and radial extension than the numerical densities from FIG \ref{fig:density_profiles012}, therefore we can ascribe the differences to the gravitational interaction between the states and the coupled system of equations that equations (\ref{eq:psi_100_2} - \ref{eq:V00_012}) represent. 
Also, we can notice the differences between an independent contribution, i.e. solving independently each state and taking the superposition of them; and a coupled contribution of each state to the total rotational curve. 
As these solutions are related with the spherical harmonics, therefore we can see that for the multi-states solutions, each one of them has a multipole contribution.
For further details about the $\ell$-boson star characteristics and within other context consult \cite{Alcubierre_2018,2021CQGra..38q4001A,2019CQGra..36u5013A}.

%%---------------------------------------------------------------------------------
\section{Data} \label{Sec:Data}
%%---------------------------------------------------------------------------------

We use a set of Low Surface Brightness Galaxies (LSBG), these type of galaxies are disk galaxies characterized by their Tully-Fisher relationship, it is similar to the high brightness surface galaxies but with low content of stars \cite{1997LSBG}, therefore it is assumed that most of their dynamics is due to the dark matter and make them good candidates to test the models.

We selected a set of 17 LSBG based on the good quality of the data classified by \cite{Data} and their mass model \cite{MassModelData,2015GuzmanRC,2018TULA} for future comparisons.  
In the table \ref{tab:characteristics_data} we showed the morphology of each galaxy in the chosen sample, for more details see \cite{Data}. 
Although their morphology could give us information about their formation history, we choose to present the results based on the radial extension and their linear behavior of the data.

\begin{table}[b]
\begin{ruledtabular}
\begin{tabular}{c|c}
\textrm{\textbf{Galaxy}}&\textrm{\textbf{Morphology}}\\
      \colrule
ESO3020120 & Spiral, hint of bar?\\ 
ESO3050090  & Barred spiral\\ 
ESO4880490  & Inclined Magellanic bar\\
UGC11557 & Fuzzy spiral, small core\\
UGC11616  & Fuzzy, irregular\\
UGC4115  & knotty and diffuse\\
ESO0140040  & Bulge, tight spiral arms\\
ESO0840411  & Edge-on\\
ESO1200211  & Fuzzy Magellanic bar\\
ESO1870510  & Irregular spiral, flocculent\\
ESO2060140  & Spiral\\
ESO4250180 & Barred open spiral\\
F730-V1 & Spiral\\
UGC11454 & Fuzzy spiral, small core\\
UGC11583 & Faint Magellanic bar\\
UGC11648 & Irregular\\
UGC11748  & Irregular, bright core/bar?\\
\end{tabular}
\end{ruledtabular}
\caption{Characteristics of each galaxy, for more details consult \cite{Data}.}
    \label{tab:characteristics_data}
\end{table}

%%---------------------------------------------------------------------------------
\section{Analysis}  \label{Sec:Analysis}
%%---------------------------------------------------------------------------------

The Bayes theorem tell us that the probability of a model $M$ with a parameter set $\theta$, given an observed data set $D$ is the posterior $\mathcal{P}$
\begin{equation}
    \mathcal{P}(\theta | D,M) = \frac{\mathcal{L}(D|\theta,M)P(\theta |M)}{E(D|M)},
\end{equation}
where $\mathcal{L}$ is  the likelihood, $P(\theta |M)$ is the prior density of the parameter vector $\theta$ for a model $M$, containing the \textit{apriori} information about the parameters of the model, and $E$ is the evidence that will be explained in section \ref{subsec:BayesFactor} due to its importance to obtain the Bayes Factor and therefore to perform the model comparison.\\
As a first step in our analysis, we calculate the maximum likelihood for the independent solutions of the SP system of equations for each state (showed in FIG \ref{fig:density_profiles01_ind}), first independently and then the superposition of each one ($\psi_{100}$, $\psi_{210}$ and $\psi_{320}$), as an approximation, being the free parameters $m_a$ and $\epsilon$ for each state. 
This analysis gave us an insight of how each state contributes to the total rotational curve, concluding that most of the galaxies in the sample must include at least three states (the ground state and two excited).\\
Because the results mentioned previously, we decided to center our work in the model described in section \ref{subsec:100,210,320} and used as priors in the Nested Sampler algorithm (NS)  the results of the maximum likelihood for that case.
As it is a coupled system of equations we cannot used the parameter $\epsilon$ of each state as a free parameter, in equation (\ref{eq:Vc2_general}) we can see that $\epsilon$ became a global parameter for this case. Consequently we choose the central amplitude of each state ($\psi_{n\ell}$(0)) as a free parameter, being part of the initial conditions needed to solved numerically the system of equations. This change made the parameter estimation more expensive computationally, since at each step in the NS the integration with the shooting method had to be made.
The free parameters for the multi-state case are: $m_{a}$, $\epsilon$, $\psi_{100}(0)$, $\psi_{210}(0)$ and $\psi_{320}(0)$.\\
  We chose the following flat priors $-26\leq\log{(m_{a}[\text{eV}/c^{2}])}\leq-20$, $-6\leq\log{(\epsilon)}\leq-2$, $-5\leq\log{(\psi_{100}(0))}\leq 0$, $-6\leq\log{(\psi_{210}(0))}\leq 0$ and $-6\leq\log{(\psi_{320}(0))}\leq 0$. For the independent cases, ground and excited states, we chose the same priors for the free parameters, $m_{a}$ and $\epsilon$, respectively. 
 For the number of live points necessary for the NS we followed the $50\times k$ rule, where $k$ corresponds to the dimensionality of the free parameter vector, as a minimum.\\
For the NS we modified the SimpleMC code that uses dynesty as an engine \cite{sergey_koposov_2022_6609296}, this sampler allowed us to obtain the Bayesian evidence which was used to obtain the Bayes factor, for more details of how the NS works consult \cite{2009Multinest}.
We used the library fgivenx \cite{fgivenx} with the output to obtain the 2$\sigma$ contours in the rotational curves of FIG \ref{fig:RC_graphs}.\\
In order to know which one of the three cases the data favours as a dark matter component, we computed the Akaike and Bayesian information criteria and the Bayes factor. The last one was compared with the ground state ($\psi_{100}$, section \ref{subsec:100}) as the based model due to its simplicity and correspondence to the soliton profile obtained from the SFDM simulations \cite{2014Schive}, usually used in the literature with an outer NFW profile \cite{2017Alma,pozo2021detection}. In addition, we take the bounds for the scalar field mass found in \cite{2018TULA} for this case, $0.212\times 10^{-23}<m_{a}[\text{eV}/c^{2}]<27.0\times 10^{-23}$. 
These criteria and the Bayes factor are defined in the following sections.
%

%%---------------------------------------------------------------------------------
\subsection{Akaike Information Criterion (AIC)}\label{subsec:AIC}
%%---------------------------------------------------------------------------------

The AIC is based on information theory and it is a way to compare models for a given data \cite{Akaike:1998zah}
\begin{equation}
    \text{AIC} =  - 2\ln{\mathcal{L}}+ 2k + \frac{2k(k+1)}{n-k-1},
\end{equation}
where $k$ is the number of fitted parameters, $n$ the number of data points and $\mathcal{L}$ is the maximum likelihood, calculated before for each model. The first term rewards the goodness of fit, while the second term penalizes the model by including extra free parameters, making it an increasing function. Therefore, the AIC discourages overfitting. By adding the last term it penalizes the fact of working with small data sets \cite{Sugiura1978}, which is our case.

%%---------------------------------------------------------------------------------
\subsection{Bayesian Information Criterion (BIC)}\label{subsec:BIC}
%%---------------------------------------------------------------------------------

Similar to the AIC, the BIC is a model selection criterion \cite{Schwarz1978EstimatingTD}
\begin{equation}
    {\rm BIC} = -2\ln{\mathcal{L}} + k\ln{n},
\end{equation}
where the first term rewards the goodness of fit, and the second term penalizes the model by including the free parameters ($k$) and the number of data points used in the fit ($n$).
%
%
%%---------------------------------------------------------------------------------
\subsection{Bayes factor}\label{subsec:BayesFactor}
%%---------------------------------------------------------------------------------

Equivalent to the information criterion, the Bayes factor allows us to compare the fitness of two models, based on the Bayes theorem.
The Bayes factor, $B_{12}$, is the ratio between the posterior of a model ($M_1$) compared to another model ($M_2$), given certain data ($D$), in logarithmic scale 
\begin{eqnarray}
    \log{B_{12}} &\equiv& \log{\left(\frac{E_{1}(D | M_1)}{E_{2}(D | M_2)}\right)} \\ &=& \log{[E_1(D|M_1)]} - \log{[E_2(D|M_2)]},
\end{eqnarray}
where $E(D|M)$ is the Bayesian evidence, defined by 
\begin{equation}
    E(D|M) = \int P(\theta |M)\mathcal{L}(D|\theta,M)d\theta.
\end{equation}
For a review about bayesian statistics and model selection see \cite{Padilla:2019mgi, 2008Trotta}.
If $\log(B_{12})$ is larger than the unity, the data slightly favours model $M_1$, if the contrary occurs ($\log(B_{12})$ is smaller than the unity), the data favours model $M_2$. The table \ref{tab:BayesFactor} contains the strength of evidence as the Jeffreys scale indicates \cite{Jeffreys61}.
\begin{table}[b]
\begin{ruledtabular}
\begin{tabular}{cc}
\textrm{\textbf{$\log{B_{12}}$}}&\textrm{\textbf{Strength}}\\
      \colrule
 $<$1.0 & Inconclusive (support $M_2$)\\ 
1.0  & Weak\\ 
2.5  & Moderate\\
5.0 & Strong\\ 
$>$5 & Very strong\\
 \end{tabular}
\end{ruledtabular}
\caption{Jeffrey's scale to quantify the Bayes factor strength.}
    \label{tab:BayesFactor}
\end{table}

%%---------------------------------------------------------------------------------
\section{Results} \label{Sec:Results}
%%---------------------------------------------------------------------------------
 
\begin{figure*}
         \centering
         \subfloat{\includegraphics[width=0.32\textwidth]{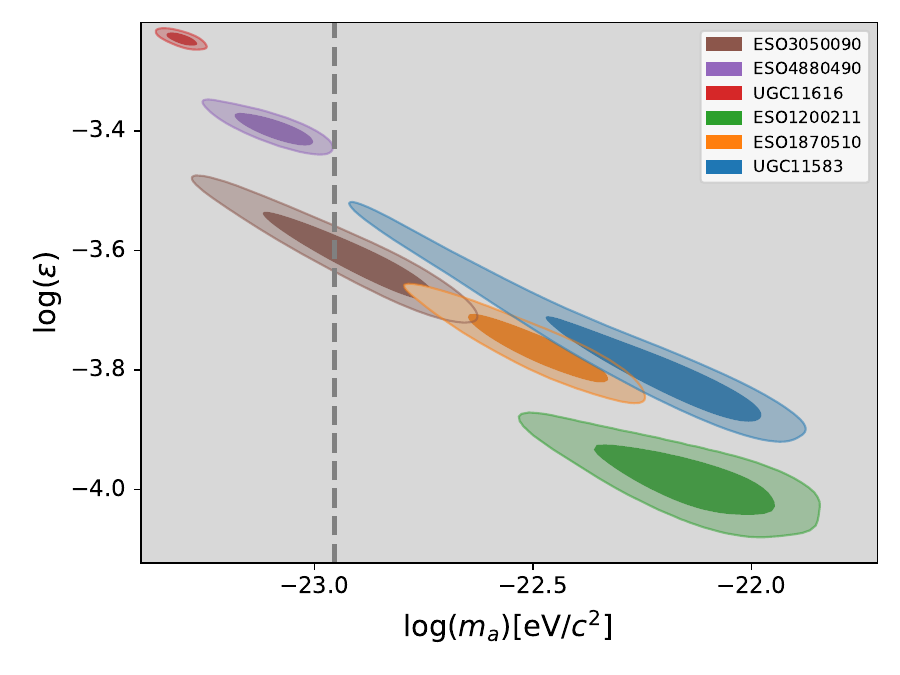}}
         \label{fig:10contours_r<10}
     \hfill
         \centering
        \subfloat{ \includegraphics[width=0.32\textwidth]{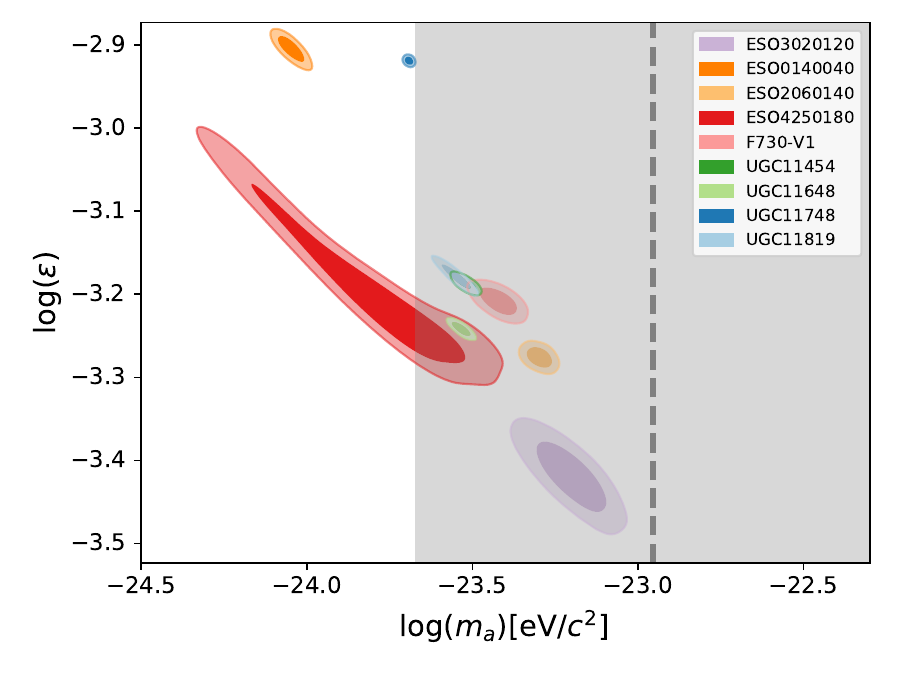}}
         \label{fig:10contours_r>10}
     \hfill
         \centering
         \subfloat{\includegraphics[width=0.32\textwidth]{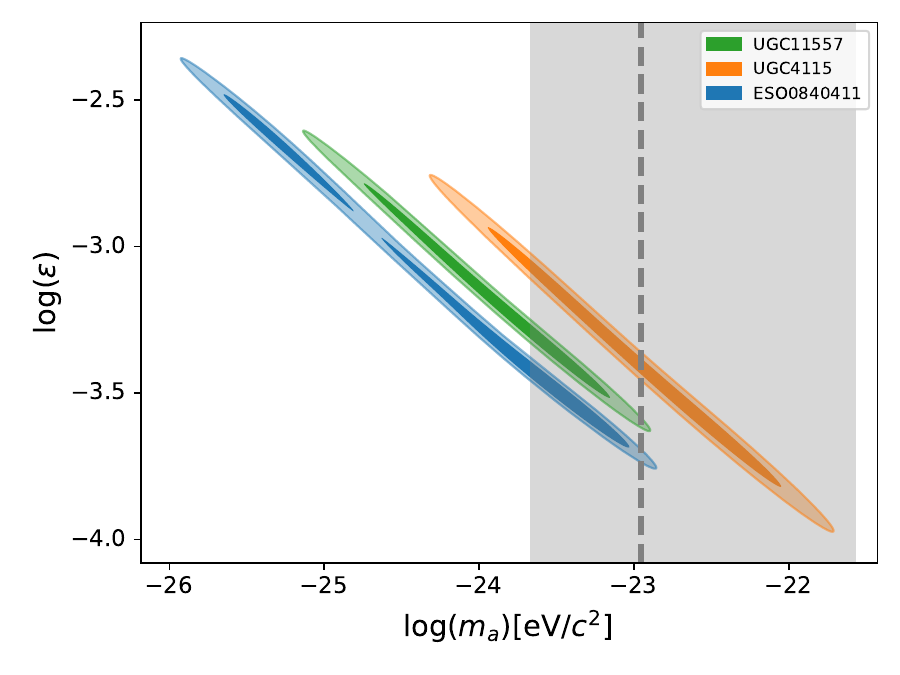}}
         \label{fig:l0contours_linear}
          
        \caption{2D marginalized posteriors distributions of the free parameters for the ground state case ($\psi_{100}$). The dashed vertical line represents $m_{a} = 1.11\times 10^{-23}$ eV/$c^{2}$ needed to have a cut-off in the power spectrum \cite{Matos_2001}. The gray band represents the bounds for the mass found in \cite{2018TULA}. Left plot contains all the galaxies with $r<10\: \textrm{kpc}$. Middle plot: all the galaxies with $r>10\: \textrm{kpc}$. Right plot: linear behavior, the three galaxies have $r<10\: \textrm{kpc}$ and they are not included in the left plot.}
        \label{fig:Contoursl0}
\end{figure*}

\begin{figure*}
         \centering
         \subfloat{\includegraphics[width=0.32\textwidth]{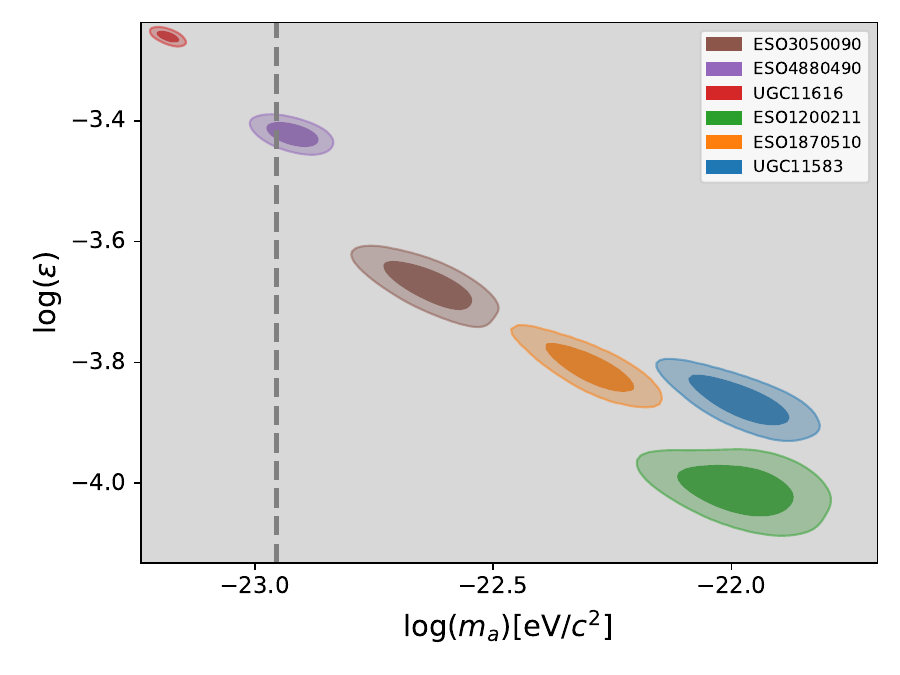}}
         \label{fig:l1contours_r<10}
     \hfill
         \centering
        \subfloat{ \includegraphics[width=0.32\textwidth]{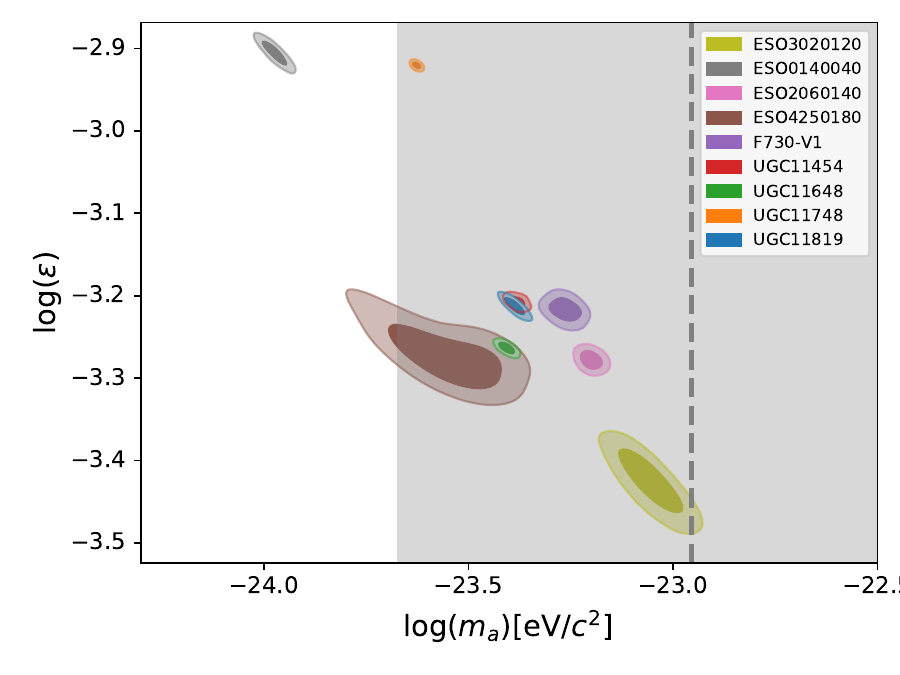}}
         \label{fig:l1contours_r>10}
     \hfill
         \centering
         \subfloat{\includegraphics[width=0.32\textwidth]{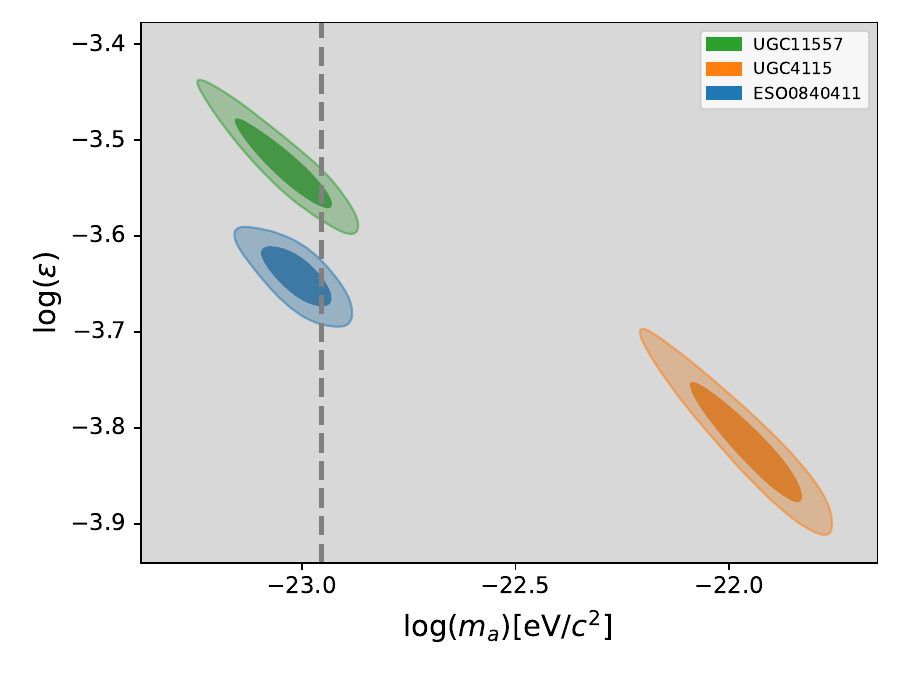}}
         \label{fig:l1contours_linear}
          
        \caption{2D marginalized posteriors distributions of the free parameters for the excited state case ($\psi_{210}$). The vertical dashed line represents the $m_{a} = 1.11\times 10^{-23}$ eV/$c^{2}$ needed to have a cut-off in the power spectrum \cite{Matos_2001}. The gray band represents the bounds for the mass found in \cite{2018TULA}. Left plot contains all the galaxies with $r<10\: \textrm{kpc}$. Middle plot: all the galaxies with $r>10\: \textrm{kpc}$. Right plot: linear behavior, the three galaxies have $r<10\: \textrm{kpc}$ and they are not included in the left plot.}
        \label{fig:Contoursl1}
\end{figure*}

Table \ref{tab:chi2_results100_210} displays the parameter estimation obtained from the NS, the $\log{(E)}$ and the $-2\ln{\mathcal{L}}$, for the cases \ref{subsec:100} and \ref{subsec:210}; the mean values are reported with $1\sigma$ confidence level. 
The $-2\ln{\mathcal{L}}$ and the $\log{(E)}$, in table \ref{tab:chi2_results100_210}, show that the cases where the contribution of only the ground state ($\psi_{100}$) is present have a better fit than the cases with only the excited state ($\psi_{210}$) for the galaxy sample. Also, it is noticeable that for all galaxies, the mass of the scalar field ($m_{a}$) is bigger for the excited state than for the ground state.\\
As mentioned in section \ref{Sec:Data}, we present our results based on the radial extension of the galaxies, they are divided in two sections: $r<10\: \textrm{kpc}$ and $r>10\: \textrm{kpc}$; adding an additional restriction for the data with a linear behavior, meaning that the data points look similar to a straight line. The shaded regions correspond to the bounds for the scalar field mass found in \cite{2018TULA}, as mentioned previously. This classification is shown in FIG \ref{fig:Contoursl0} and \ref{fig:Contoursl1}, for the independent cases, $\psi_{100}$ and $\psi_{210}$, respectively. 
Here we notice that most of the galaxies with $r<10\: \textrm{kpc}$ tend to have bigger masses while the galaxies with $r>10\: \textrm{kpc}$ tend to prefer lighter masses. An important case raises in these contour plots, the galaxy UGC11616 has a radial extension of $r=9.6\: \textrm{kpc}$, due to the proximity to the $r=10\: \textrm{kpc}$ value, it follows the $r>10\: \textrm{kpc}$ behavior.
On the other hand, the galaxy data with linear behavior and $r<10\: \textrm{kpc}$ are highly correlated on the free parameters in the ground state case ($\psi_{100}$), this correlation seems to be broken on the excited state case ($\psi_{210}$).

\begin{figure*}
         \centering
         \subfloat{\includegraphics[width=0.32\textwidth]{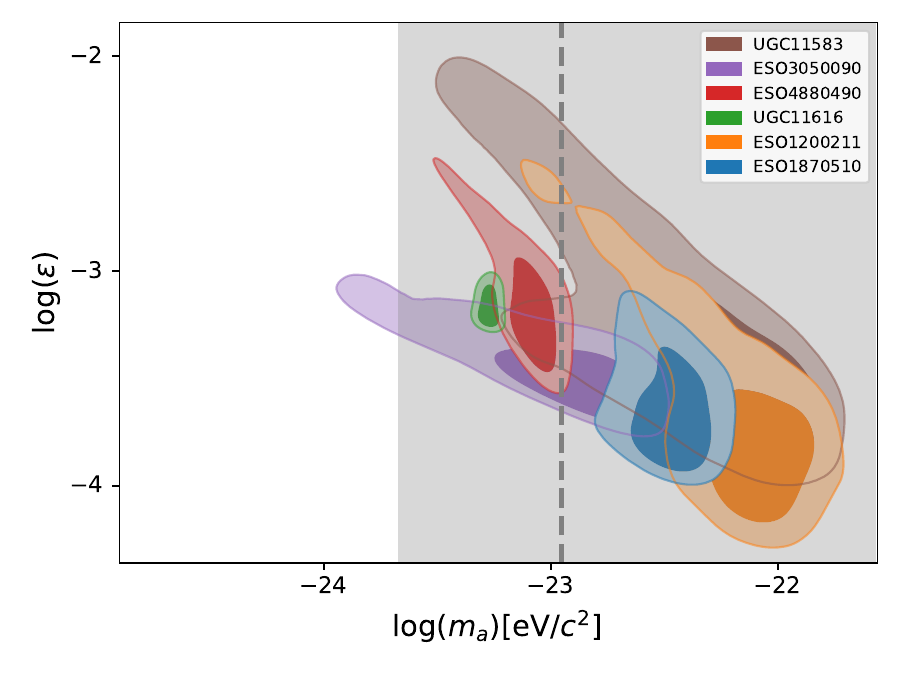}}
         \label{fig:l012_coupledcontours_r<10}
     \hfill
         \centering
        \subfloat{ \includegraphics[width=0.32\textwidth]{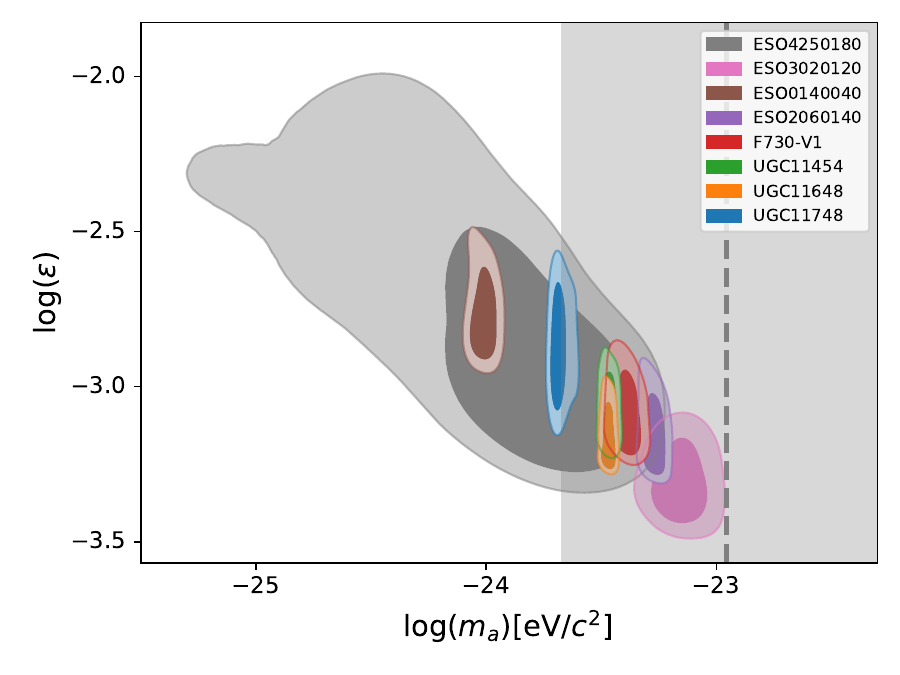}}
         \label{fig:l012_coupled_contours_r>10}
     \hfill
         \subfloat{\includegraphics[width=0.32\textwidth]{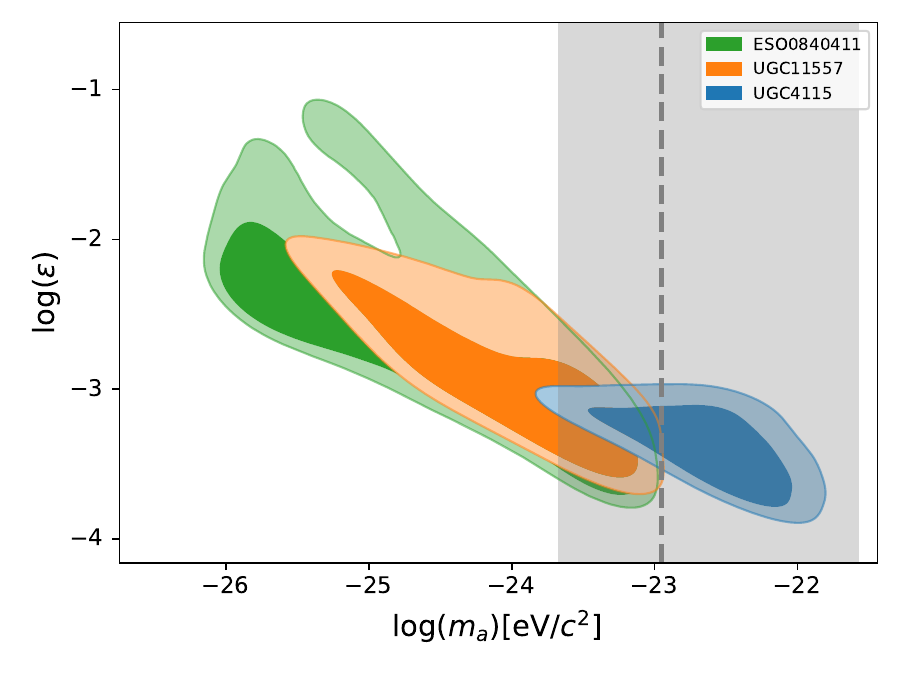}}
         \label{fig:l012_coupled_contours_linear}
          
        \caption{2D marginalized posteriors distributions of the free parameters for the multi-state case ($\psi_{100}$, $\psi_{210}$ and $\psi_{320}$). The vertical dashed line represents the $m_{a} = 1.11\times 10^{-23}$ eV/$c^{2}$ needed to have a cut-off in the power spectrum \cite{Matos_2001}. The gray band represents the bounds for the mass found in \cite{2018TULA}. Left plot contains all the galaxies with $r<\: \textrm{kpc}$. Middle plot: all the galaxies with $r>10\: \textrm{kpc}$. Right plot: linear behavior, the three galaxies have $r<10\: \textrm{kpc}$ and they are not included in the left plot.}
        \label{fig:Contoursl012}
\end{figure*}

\begin{figure}
    \centering
    \includegraphics[width=2.8in]{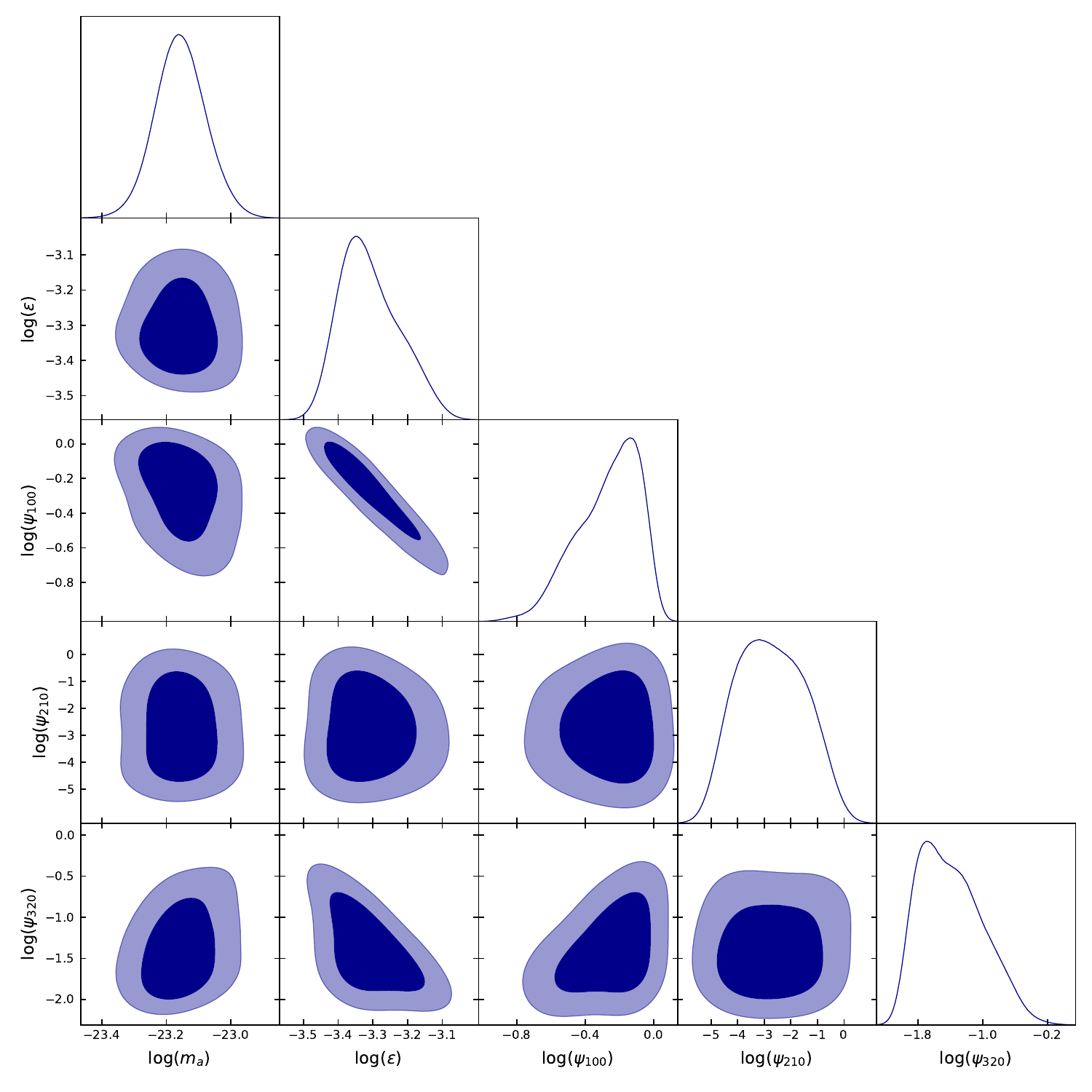}
    \caption{Triangle plot for the galaxy ESO3020120.}
    \label{fig:Triangle_ESO3020120}
\end{figure}

The table \ref{tab:nested_results_multistate} contains the parameter constraints, the $\log{(E)}$ and the $-2\ln{\mathcal{L}}$ for the multi-state case (\ref{subsec:100,210,320}).  The mean values are reported along with with the $1\sigma$ confidence level. The contour plots related with these results are shown in FIG \ref{fig:Contoursl012}, where we have followed the classification mentioned before. 
The galaxies followed the same trend as the independent cases, those with $r<10\: \textrm{kpc}$ tended to have bigger masses while the galaxies with $r>10\: \textrm{kpc}$ tended to prefer lighter masses. For the galaxies with linear behavior the correlation seemed to be diminished between $m_{a}$ and $\epsilon$. 
Although, for all galaxies, the correlation still remains between $m_{a}$ and  the central amplitude of the first state ($\psi_{100}$(0)), as we can see in FIG \ref{fig:Triangle_ESO3020120}.
To see the seventeen triangle plots for this case and the independent cases, see the  
\href{https://github.com/atalianb/Triangle_plots_ell_boson_stars}{repository}\footnote{\url{https://github.com/atalianb/Triangle_plots_ell_boson_stars}}.

\begin{figure*}
     \centering
     \begin{subfigure}[b]{0.26\textwidth}
         \centering
         \includegraphics[width=\textwidth]{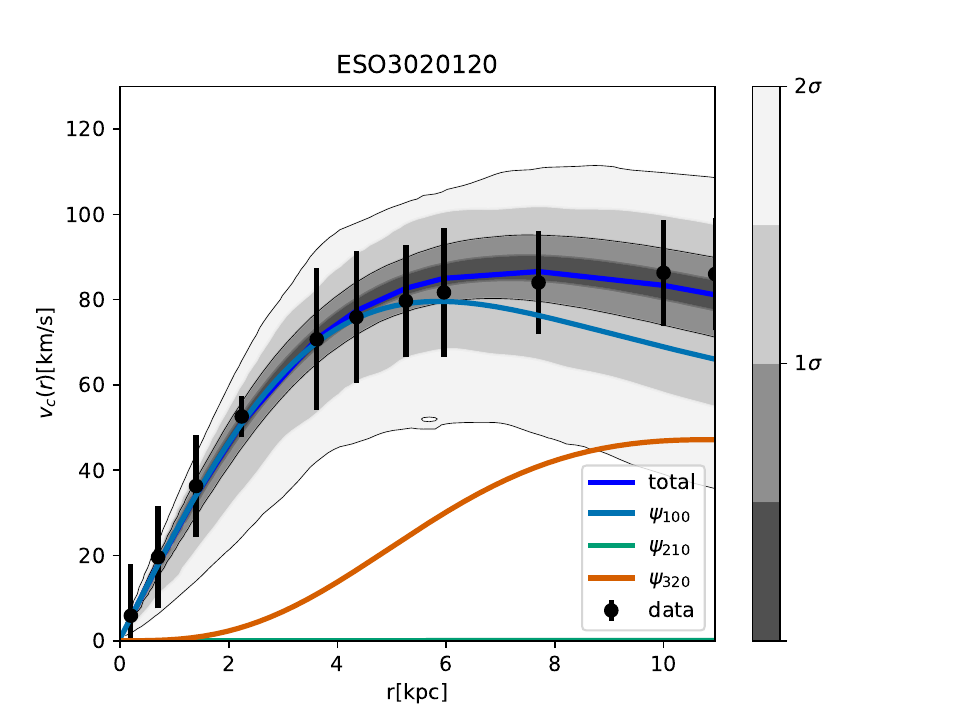}
         \label{fig:ESO3020120_rotcurv}
     \end{subfigure}
     \begin{subfigure}[b]{0.26\textwidth}
         \centering
         \includegraphics[width=\textwidth]{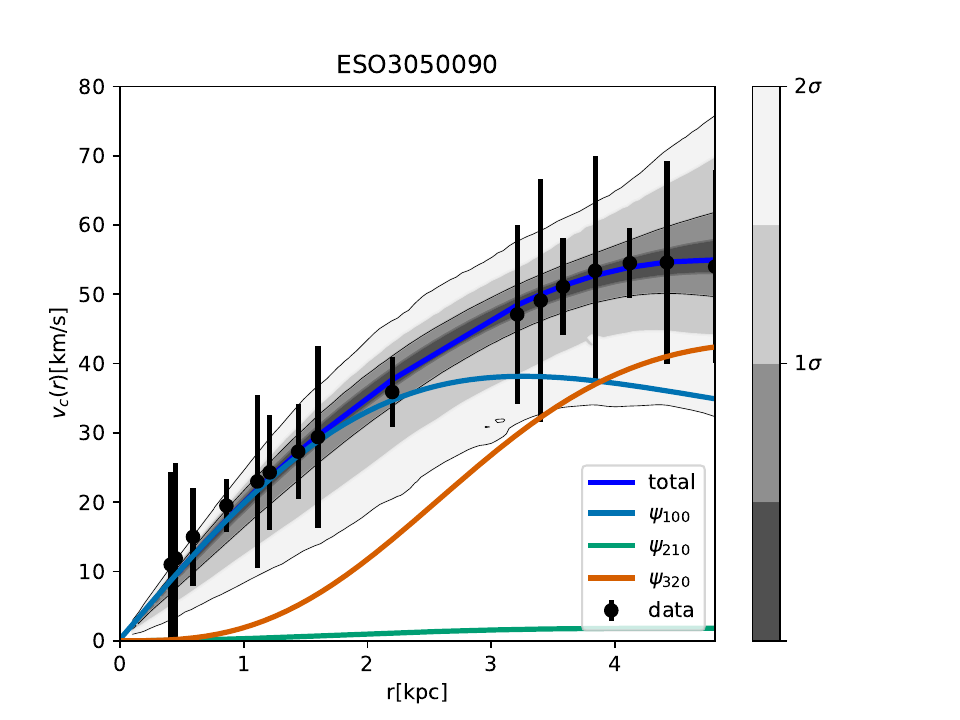}
         \label{fig:ESO3050090_rotcurv}
     \end{subfigure}
     \begin{subfigure}[b]{0.26\textwidth}
         \centering
         \includegraphics[width=\textwidth]{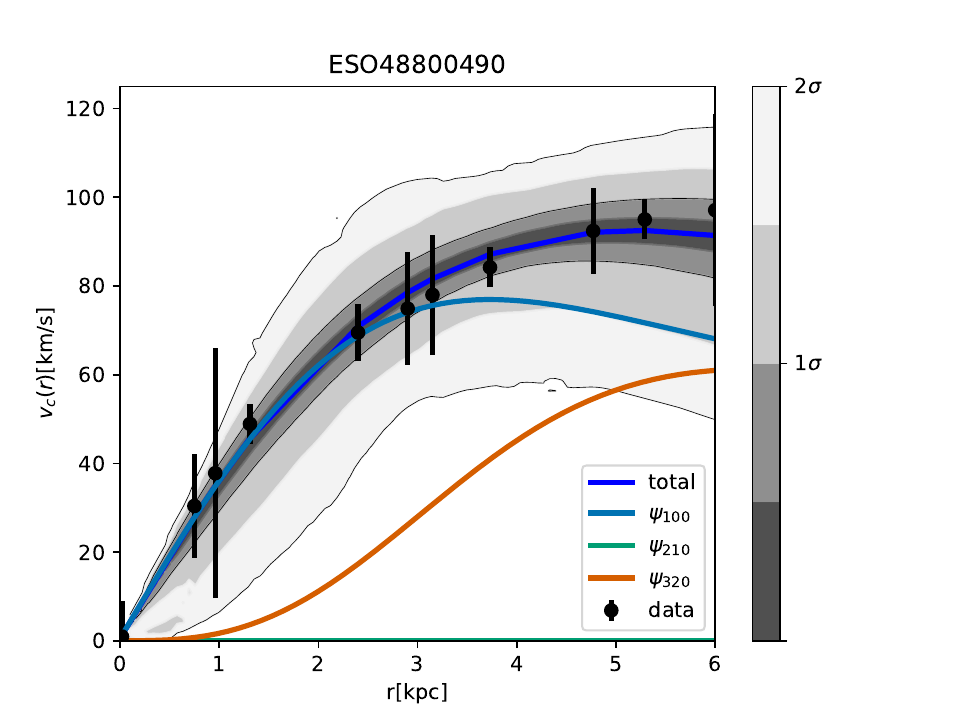}
         \label{fig:ESO48800490_rotcurv}
     \end{subfigure}
     \begin{subfigure}[b]{0.26\textwidth}
         \centering
         \includegraphics[width=\textwidth]{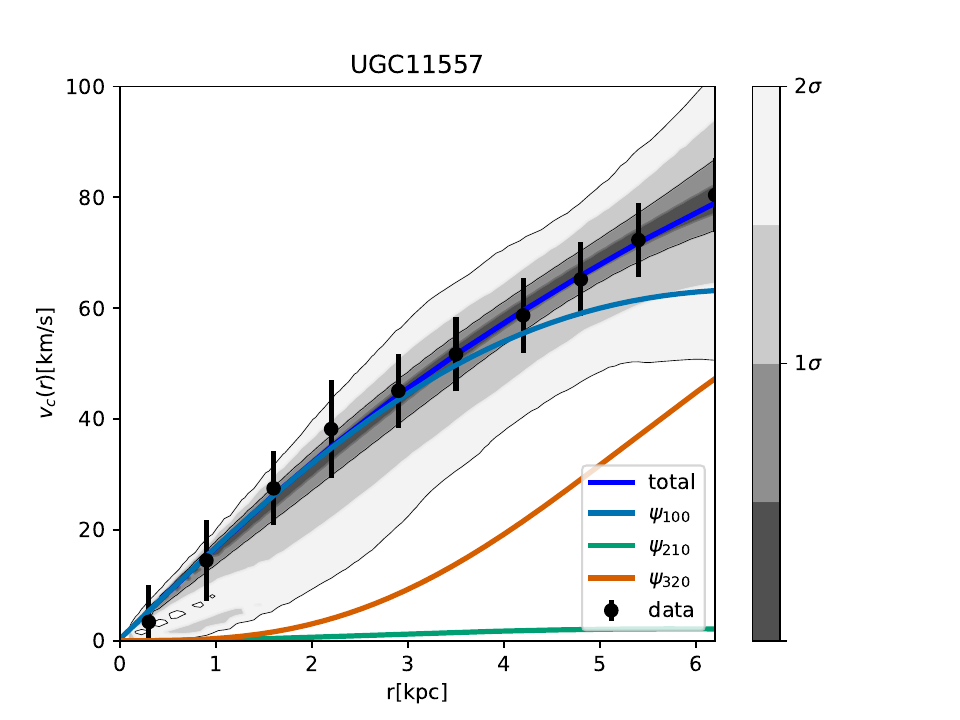}
         \label{fig:UGC11557_rotcurv}
     \end{subfigure}
     \begin{subfigure}[b]{0.26\textwidth}
         \centering
         \includegraphics[width=\textwidth]{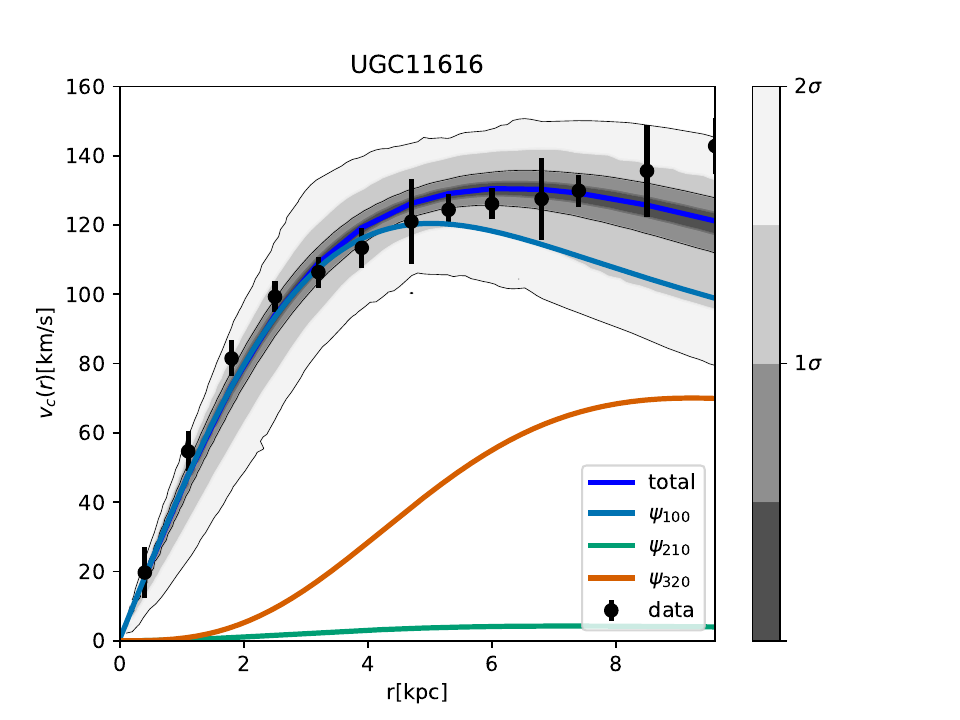}
         \label{fig:UGC11616_rotcurv}
     \end{subfigure}
     \begin{subfigure}[b]{0.26\textwidth}
         \centering
         \includegraphics[width=\textwidth]{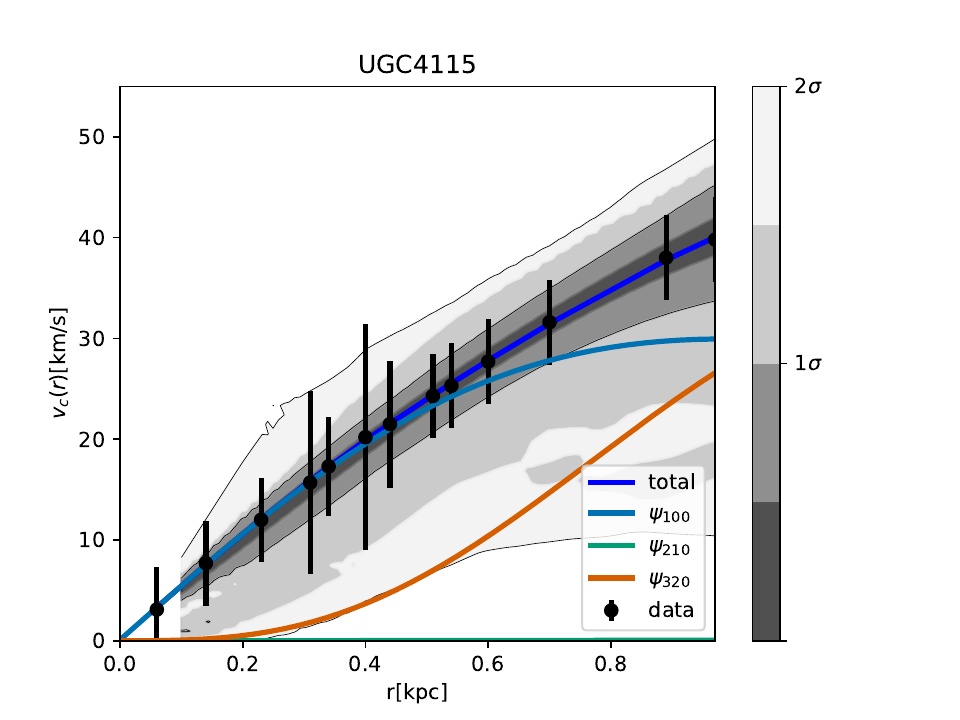}
         \label{fig:U4115_rotcurv}
     \end{subfigure}
     \begin{subfigure}[b]{0.26\textwidth}
         \centering
         \includegraphics[width=\textwidth]{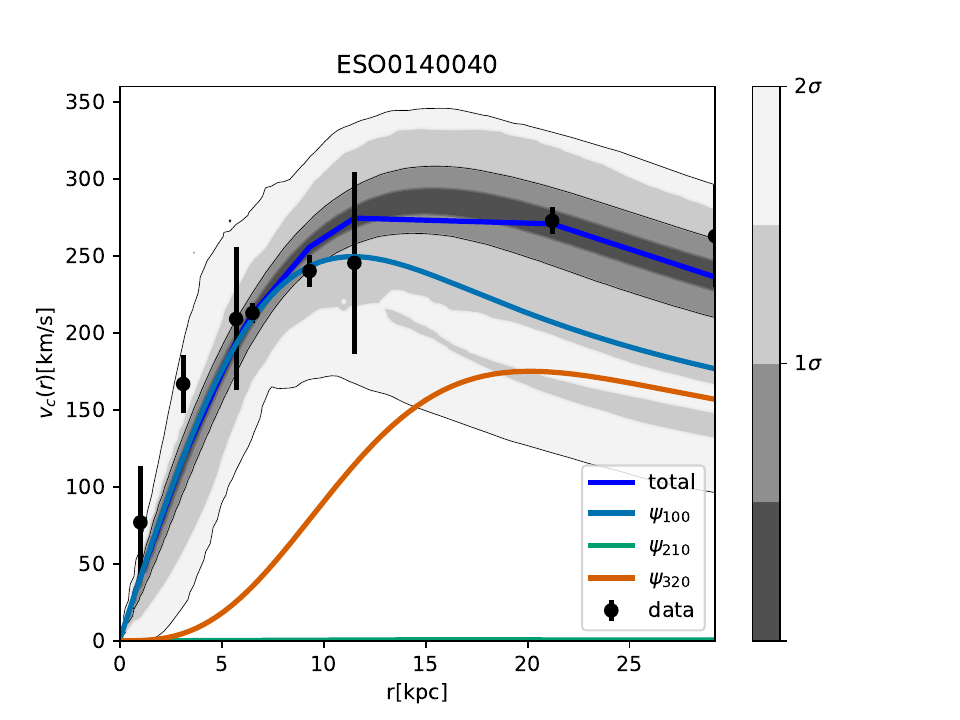}
         \label{fig:ESO0140040_rotcurv}
     \end{subfigure}
     \begin{subfigure}[b]{0.26\textwidth}
         \centering
         \includegraphics[width=\textwidth]{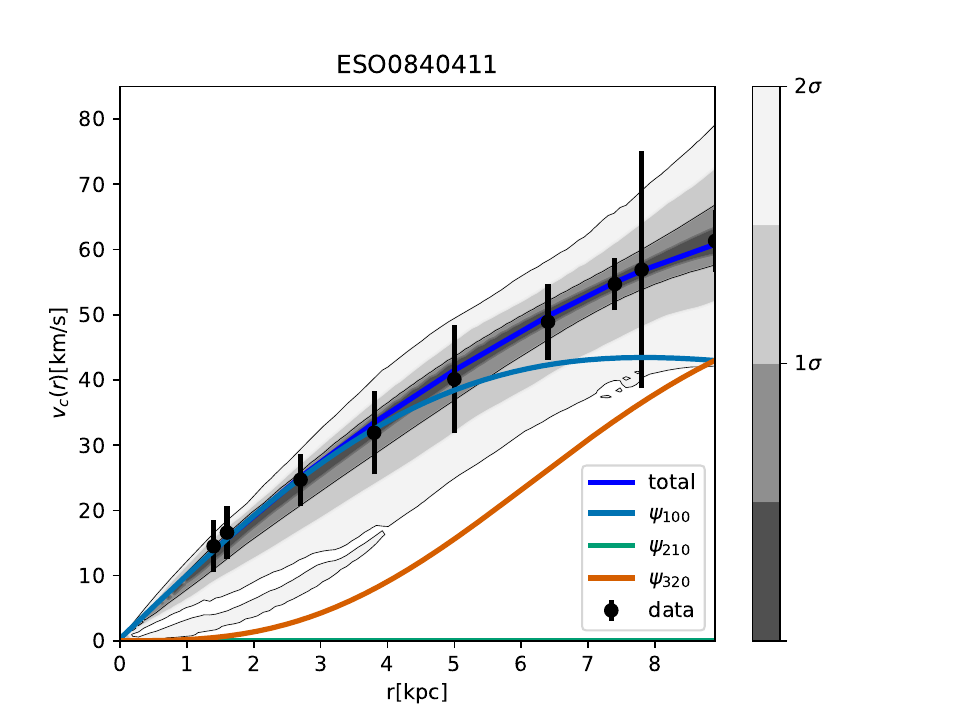}
         \label{fig:ESO0840411_rotcurv}
     \end{subfigure}
     \begin{subfigure}[b]{0.26\textwidth}
         \centering
         \includegraphics[width=\textwidth]{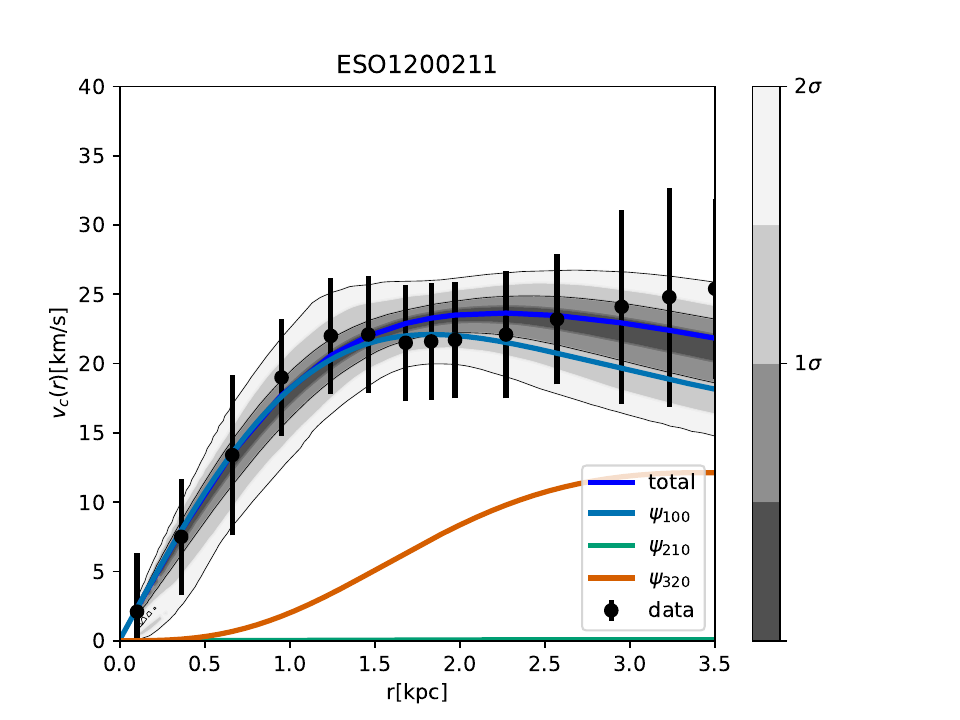}
         \label{fig:ESO1200211_rotcurv}
     \end{subfigure}
     \hfill
     \begin{subfigure}[b]{0.26\textwidth}
         \centering
         \includegraphics[width=\textwidth]{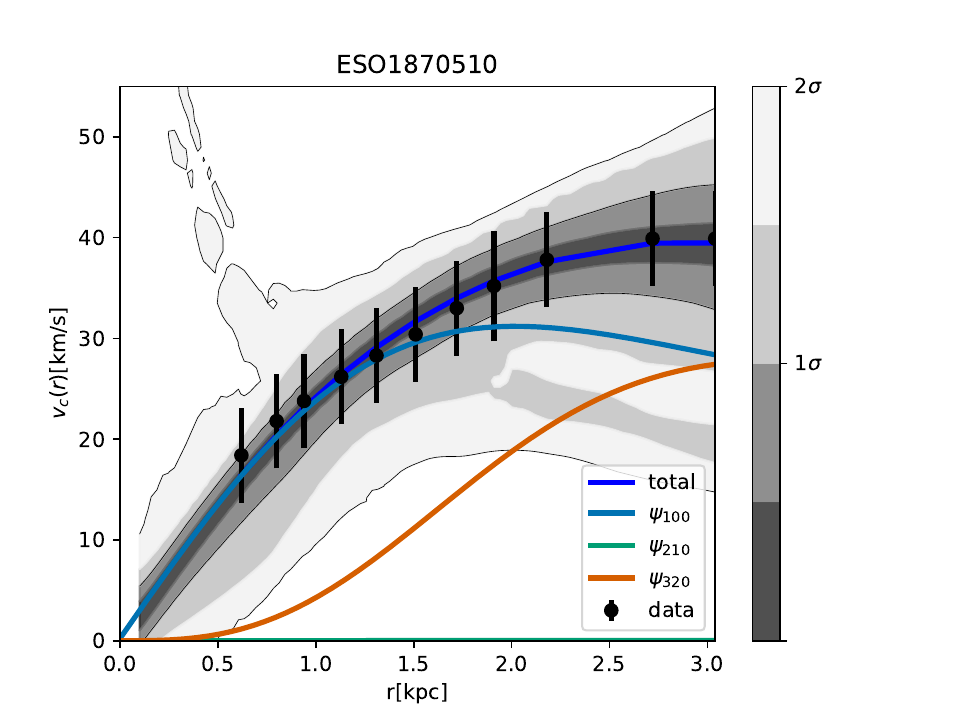}
         \label{fig:ESO1870510_rotcurv}
     \end{subfigure}
     \begin{subfigure}[b]{0.26\textwidth}
         \centering
         \includegraphics[width=\textwidth]{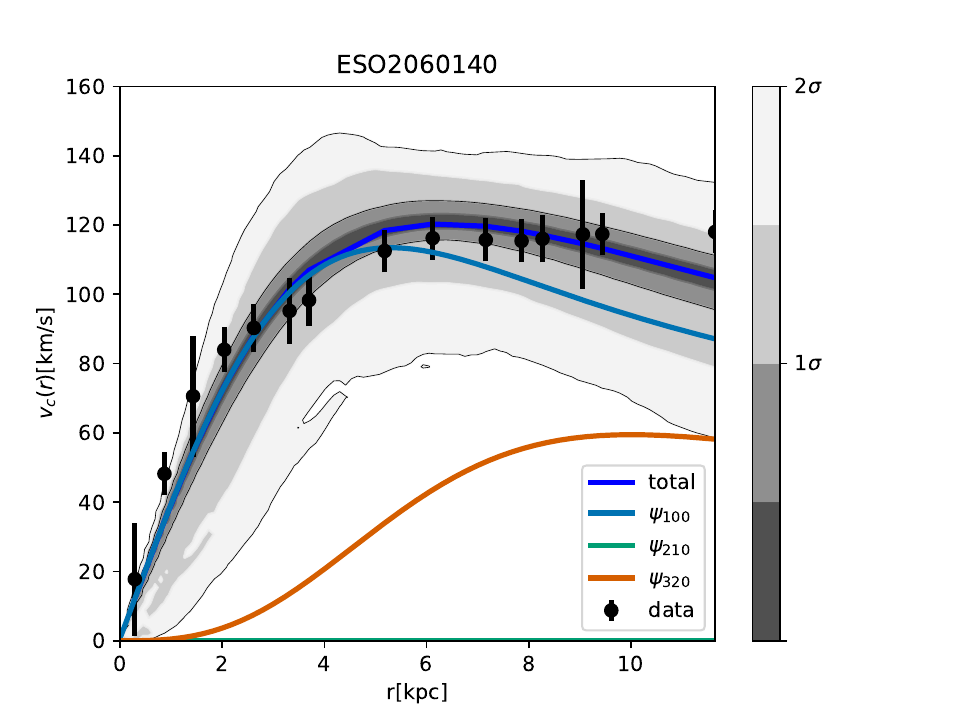}
         \label{fig:ESO2060140_rotcurv}
     \end{subfigure}
     \begin{subfigure}[b]{0.26\textwidth}
         \centering
         \includegraphics[width=\textwidth]{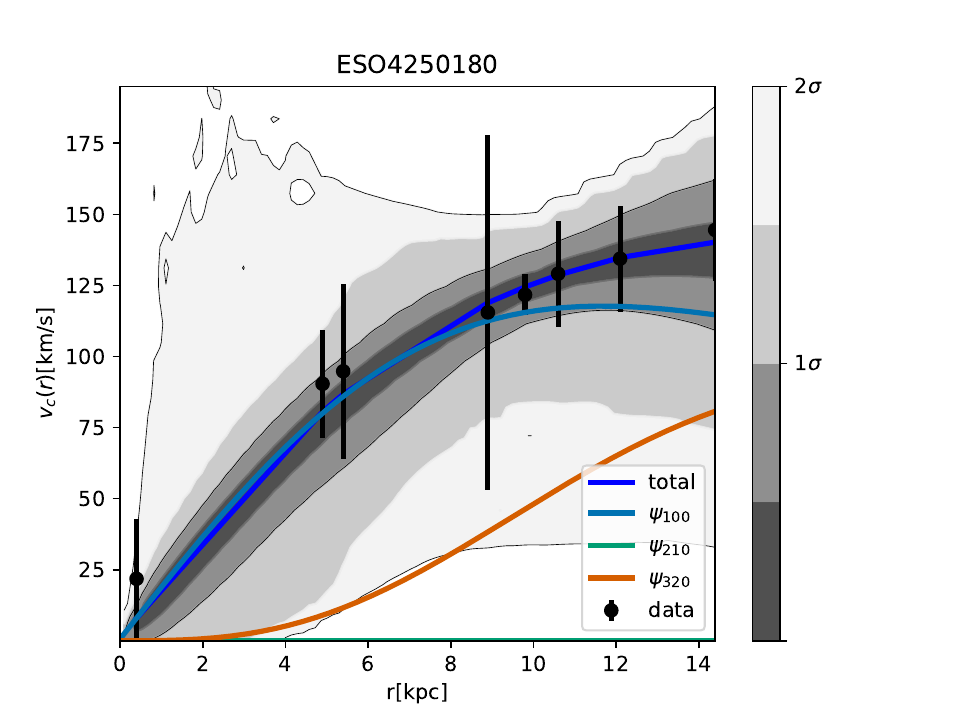}
         \label{fig:ESO4250180_rotcurv}
     \end{subfigure}
     \begin{subfigure}[b]{0.26\textwidth}
         \centering
         \includegraphics[width=\textwidth]{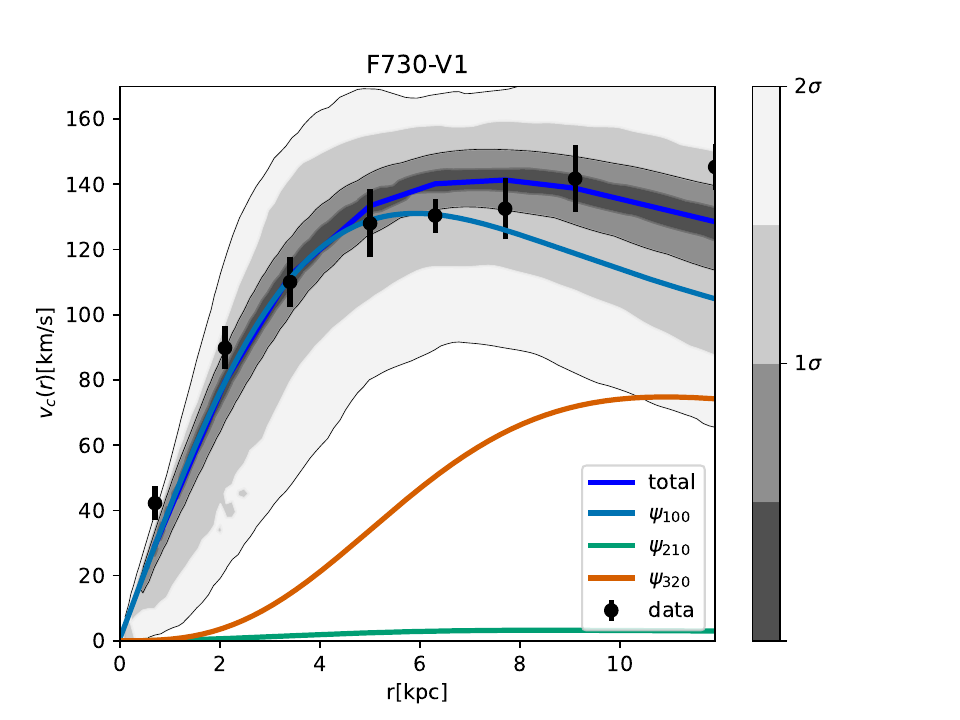}
         \label{fig:F730-V1_rotcurv}
     \end{subfigure}
     \begin{subfigure}[b]{0.26\textwidth}
         \centering
         \includegraphics[width=\textwidth]{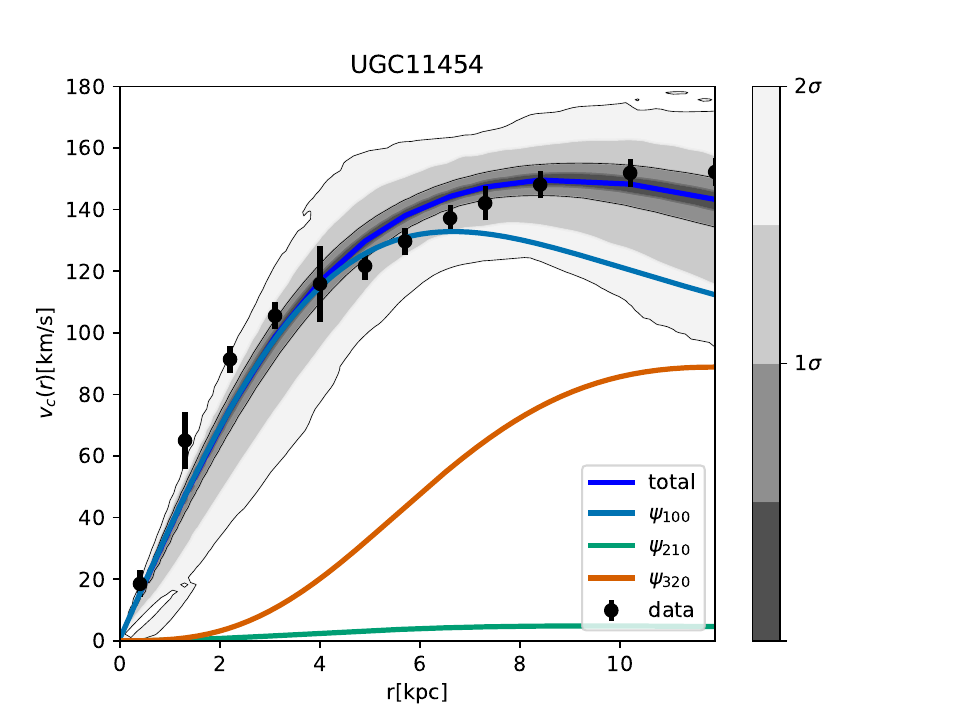}
         \label{fig:UGC11454_rotcurv}
     \end{subfigure}
     \begin{subfigure}[b]{0.26\textwidth}
         \centering
         \includegraphics[width=\textwidth]{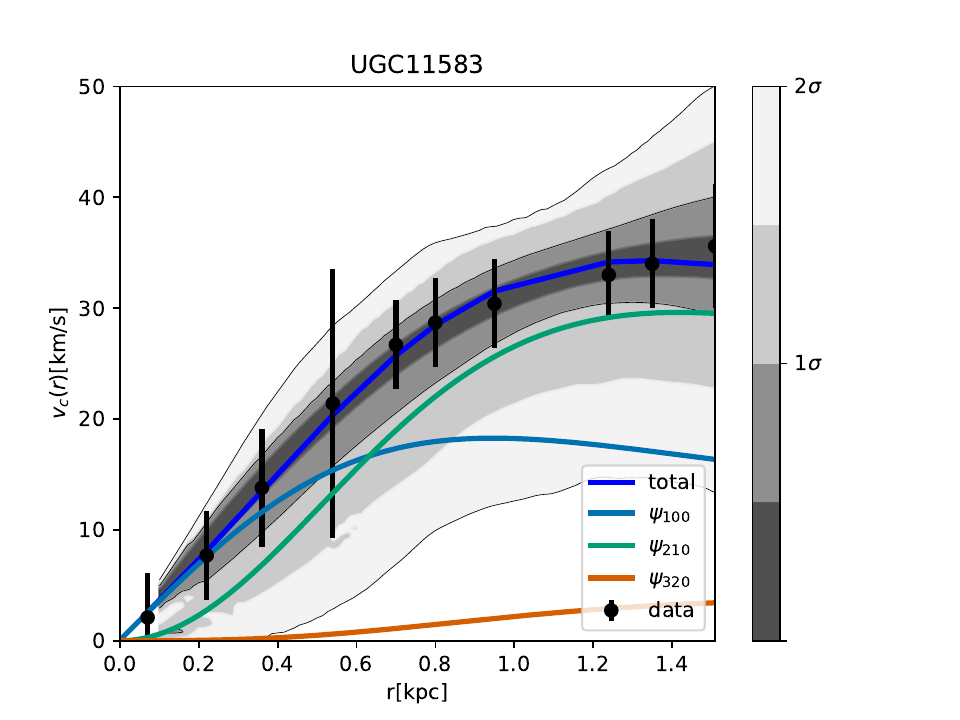}
         \label{fig:UGC11583_rotcurv}
     \end{subfigure}
     \begin{subfigure}[b]{0.26\textwidth}
         \centering
         \includegraphics[width=\textwidth]{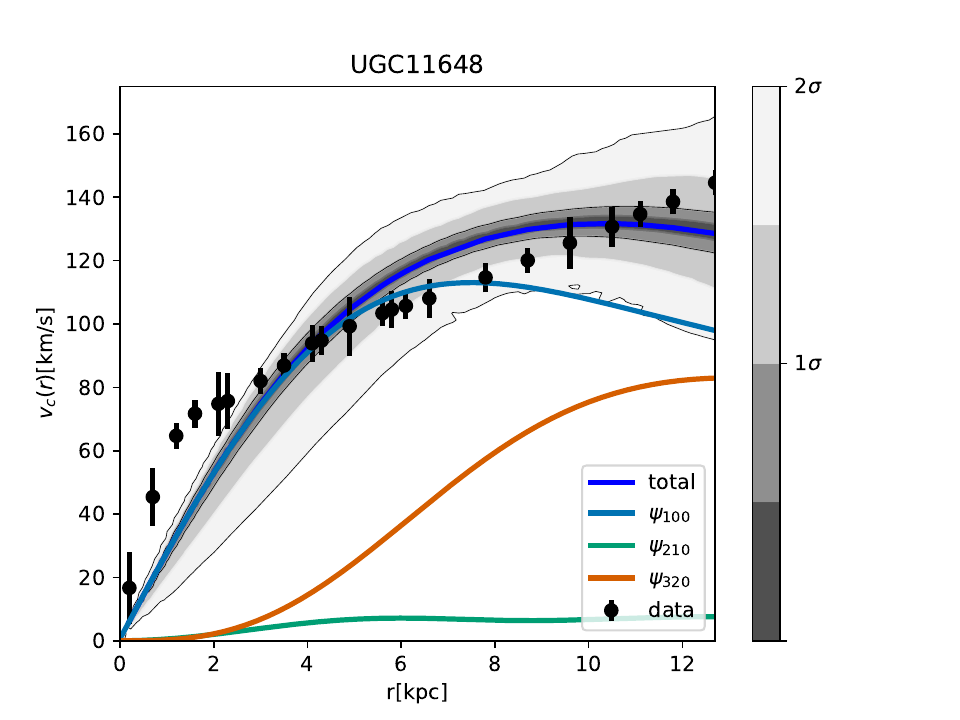}
         \label{fig:UGC11648_rotcurv}
     \end{subfigure}
     \begin{subfigure}[b]{0.26\textwidth}
         \centering
         \includegraphics[width=\textwidth]{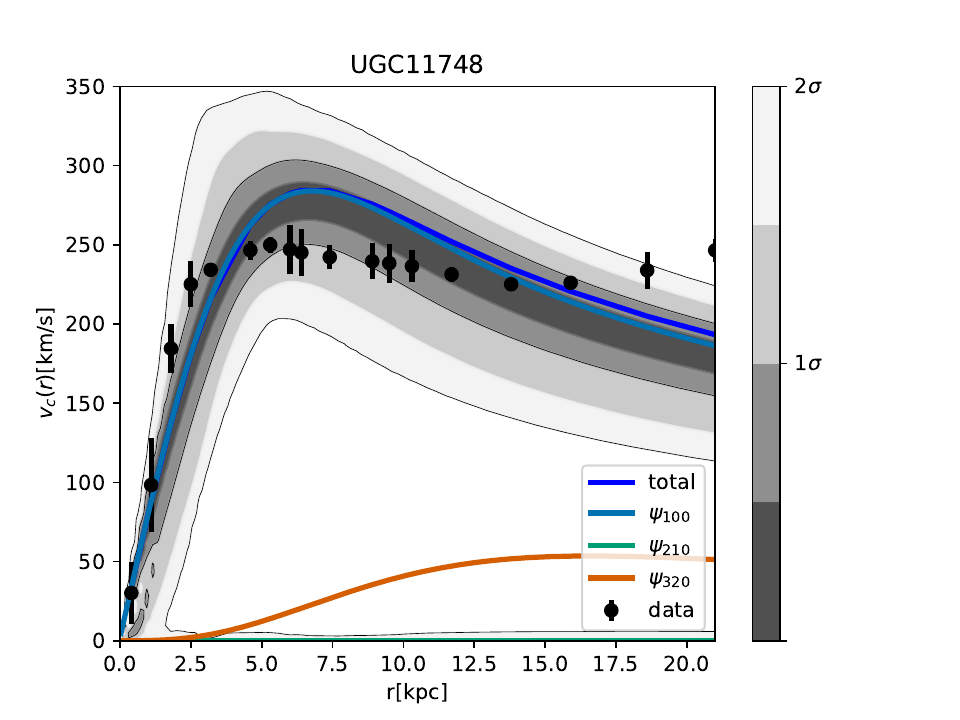}
         \label{fig:UGC11748_rotcurv}
     \end{subfigure}
        \caption{Rotational curve for each galaxy with the parameter estimation obtained from the NS. The contour lines indicate the accuracy of the parameter estimation at 1$\sigma$ and 2$\sigma$ as the gray color bar shows. The blue dark line indicates the resulting rotational curve, the blue light line indicates the contribution of the state $\psi_{100}$, the green the contribution of the state $\psi_{210}$ and the orange the contribution of $\psi_{320}$.}
        \label{fig:RC_graphs}
\end{figure*}

The plots in FIG \ref{fig:RC_graphs} show the parameter estimation results reported in table \ref{tab:nested_results_multistate} at 1$\sigma$ and 2$\sigma$ as the gray color bar shows. We can observe that most galaxies have an interesting behavior regarding the contributions from each state to the total rotational curve (dark blue line), where the blue line corresponds to the ground state ($\psi_{100}$) shows a predominant contribution, $\psi_{210}$ (green line) contributes less than $\psi_{320}$ (orange line). 
Both of them contribute to the larger radius while the ground state ($\psi_{100}$) remains in the center, suggesting that the $\psi_{210}$ contribution is closer to zero. One can notice that galaxy UGC11583, one of the smallest with a $r=1.5 \: \textrm{kpc}$ radial extension has a different contribution from each state, where $\psi_{210}$ is the predominant one as $\psi_{320}$ has a smaller amplitude.
 Although we obtain an acceptable parameter inference for all the data sample according to the gray contours, we can observe that rotational curves for galaxies UGC11648 and UGC11748 the multi-state case does not fit the data too well. 
It is important to mention that unlike the NS results for the independent cases where the convergence is clear, in the multi-state case, specifically for the central amplitudes of each state the convergence is not so clear and particularly the ground state central amplitude ($\psi_{100}(0)$) seems to have a boundary that corresponds to the prior upper limit, $\log{\psi_{100}(0)}=0$.
 The results for the NS used to obtain the plots and the plots themselves can be consulted in the \href{https://github.com/atalianb/fgivenx_plots}{repository}\footnote{\url{https://github.com/atalianb/fgivenx_plots}}.
 
\begin{figure*}
    \centering
    \includegraphics[width=6in]{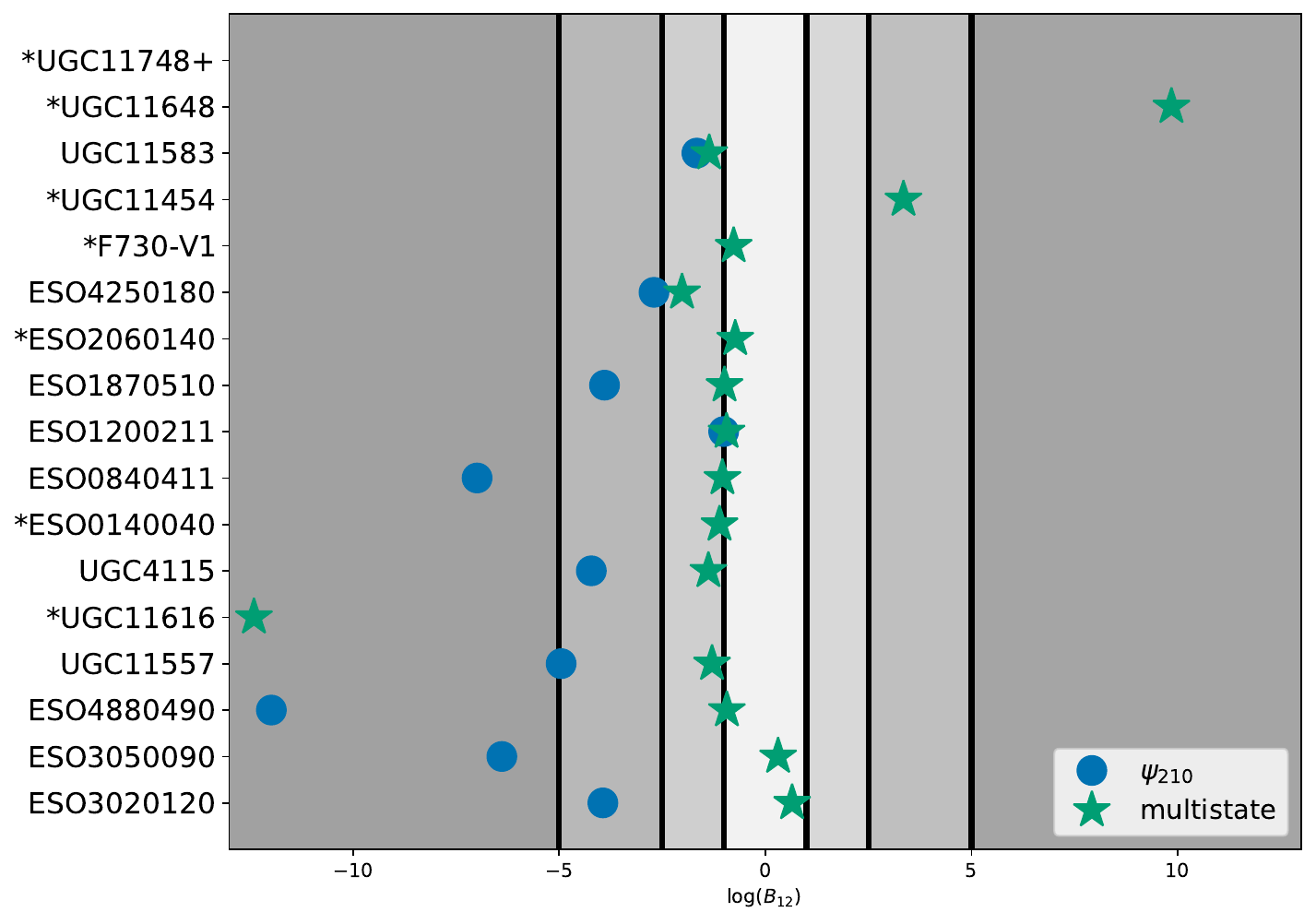}
    \caption{Bayes Factor ($\log(B_{12})$) for the comparisons between the ground state case ($\psi_{100}$) as the base model and the other two cases. The blue circles correspond to the excited state case ($\psi_{320}$) and the green stars to the multi-state case. The shaded regions indicate the evidence strength and the  solid lines represent the values mentioned in table \ref{tab:BayesFactor}. Galaxies marked with a * have a smaller Bayes factor value and galaxies with a + have a bigger value, those can be consulted in table \ref{tab:model_comparison}.}
    \label{fig:BayesFactor}
\end{figure*}

In table \ref{tab:model_comparison}, the AIC, BIC, $\log(B_{12})$ and $-2\ln{\mathcal{L}}$ are reported  for each case. The AIC and BIC values for the three cases are similar with a noticeable decrease for the ground state. 
As for the $-2\ln{\mathcal{L}}$ value, it is noticed that for all galaxies is bigger for the excited state ($\psi_{210}$) and the multi-state case smaller than the ground state, except for galaxy UGC11648 with a small increase. 
Particularly, the Bayes factor for each galaxy is represented in FIG \ref{fig:BayesFactor}, where the shaded regions correspond to the strength in Jeffrey's scale mentioned in table \ref{tab:BayesFactor}. Galaxies with an asterisk ($\ast$) have smaller $\log{(B_{12})}$ for the $\psi_{210}$ case and those (UGC11748) with a plus marker ($+$) have bigger $\log{(B_{12})}$ for the multi-state case, that doesn't appear in the figure and can be consulted in table \ref{tab:model_comparison}. 
The position of the blue dots indicate that most galaxies prefer the ground state rather than the excited state $\psi_{210}$, except for UGC11583 and ESO1200211 that seem to slightly prefer the excited state. 
On the other hand, the green stars position suggest that the data moderately favours the multi-state case even though the number of free parameters is bigger compare to the ground state. Some galaxies stand out, UGC11616 with a $\log{(B_{12})}=-12.40$ that supports the ground state and, galaxies UGC11648 and UGC11454 strongly supporting the multi-state case.

\begin{table*}
\begin{ruledtabular}
\begin{tabular}{c|cccc|cccc}
         & \multicolumn{4}{c|}{\textbf{$\psi_{100}$}}&\multicolumn{4}{c}{\textbf{$\psi_{210}$}}\\
         \hline
 \textbf{Galaxy} &$\log{(m_{a})}$ & $\log{(\epsilon)}$ & $\log{(E)}$ & \textbf{$-2\log{\mathcal{L}}$} &$\log{(m_{a})}$ & $\log{(\epsilon)}$ & $\log(E)$ & \textbf{$-2\ln{\mathcal{L}}$}\\ 
\hline
 ESO3020120 &   -23.20$^{+0.08}_{-0.10}$ & -3.42$^{+0.03}_{-0.04}$  & -7.93$\pm$ 0.36 & 0.96 &  -23.05$^{+0.07}_{-0.07}$ & -3.42$^{+0.03}_{-0.04}$  & -11.87 $\pm$ 0.36 & 8.61\\ 
ESO3050090 & -22.88$^{+0.15}_{-0.25}$ & -3.62$^{+0.09}_{-0.05}$ & -7.05 $\pm$ 0.32 & 0.98 & -22.63 $^{+0.03}_{-0.04}$ & -3.67$^{+0.04}_{-0.04}$  & -13.44 $\pm$ 0.37 & 11.11\\ 
ESO4880490 & -23.09$^{+0.08}_{-0.09}$ & $-3.40^{+0.03}_{-0.02}$ & -8.89 $\pm$ 0.36 & 2.8 & -22.92$^{+0.05}_{-0.05}$ & -3.42$^{+0.02}_{-0.02}$ & -20.87 $\pm$ 0.39 & 24.7\\
UGC11557 &  -23.45$^{+0.25}_{-1.49}$ & -3.39$^{+0.71}_{-0.10}$ & -6.61 $\pm$  0.33 & 0.52 & -23.03$^{+0.08}_{-0.22}$ &  -3.53$^{+0.05}_{-0.04}$& -11.35 $\pm$ 0.36& 7.61\\
UGC11616 & -23.31$^{+0.03}_{-0.03}$ & -3.25$^{+0.01}_{-0.01}$ & -21.02 $\pm$ 0.41 & 23.44 & -23.18$^{+0.02}_{-0.02}$ &  -3.26$^{+0.01}_{-0.01}$ & -62.66 $\pm$ 0.417 & 105.97\\
UGC4115 & -22.39$^{+0.29}_{-1.60}$& -3.67$^{+0.76}_{-0.12}$& -5.75 $\pm$ 0.31 & 0.04 &  -21.95$^{+0.11}_{-0.13}$ & -3.82$^{+0.06}_{-0.05}$ & -9.97 $\pm$ 0.36 & 4.77\\
ESO0140040 & -24.05$^{+0.04}_{-0.03}$ & -2.90$^{+0.01}_{-0.01}$ & -15.62 $\pm$ 0.39 & 13.36 & -23.98$^{+0.03}_{-0.03}$ & -2.90$^{+0.01}_{-0.01}$ &-38.07 $\pm$ 0.41 & 57.27\\
ESO0840411 & -23.29$^{+0.20}_{-2.20}$ & -3.54$^{+1.04}_{-0.08}$ & -6.480 $\pm$ 0.320 & 0.45 & -23.00$^{+0.06}_{-0.08}$ & -3.64$^{+0.03}_{-0.03}$ & -13.47 $\pm$ 0.36 & 12.21\\
ESO1200211 & -22.12$^{+0.17}_{-0.21}$ & -3.99$^{+0.06}_{-0.05}$ & -7.57 $\pm$ 0.34 & 1.56 & -21.98$^{+0.10}_{-0.11}$ & -4.01$^{+0.04}_{-0.04}$ & -8.58 $\pm$ 0.33 & 4.51\\
ESO1870510 & -22.47$^{+0.12}_{-0.16}$ & -3.77$^{+0.06}_{-0.05}$ & -7.76 $\pm$ 0.35 & 0.82 & -22.30$^{+0.09}_{-0.09}$ & -3.81$^{+0.04}_{-0.04}$ & -11.66 $\pm$ 0.36 & 8.16\\
ESO2060140 & -23.30$^{+0.04}_{-0.03}$ & -3.28$^{+0.01}_{-0.01}$ & -21.31 $\pm$ 0.42 & 23.36 & -23.20$^{+0.03}_{-0.02}$ & -3.28$^{+0.01}_{-0.01}$ & -54.49 $\pm$ 0.42 & 89.96\\
ESO4250180 & -23.72$^{+0.21}_{-0.41}$ & -3.23$^{+0.15}_{-0.05}$ & -6.88 $\pm$ 0.32 & 1.25 & -23.53$^{+0.12}_{-0.17}$ & -3.28$^{+0.05}_{-0.03}$ & -9.58 $\pm$ 0.35 & 30.26\\
F730-V1 & -23.42$^{+0.05}_{-0.05}$ & -3.21$^{+0.01}_{-0.01}$ & -21.79 $\pm$ 0.28 & 25.36 & -23.26$^{+0.03}_{-0.04}$ & -3.22$^{+0.02}_{-0.02}$ & -54.97 $\pm$ 0.4 & 91.74\\
UGC11454 & -23.52$^{+0.03}_{-0.03}$ & -3.19$^{+0.01}_{-0.01}$ & -35.45 $\pm$ 0.31 & 49.97 & -23.28$^{+0.02}_{-0.02}$ & -3.21$^{+0.01}_{-0.01}$ & -128.28 $\pm$ 0.43 & 236.57\\
UGC11583 & -22.16$^{+0.18}_{-0.31}$ & -3.82$^{+0.11}_{-0.06}$ & -6.57 $\pm$ 0.32 & 0.5 & -21.98$^{+0.09}_{-0.10}$ & -3.86$^{+0.04}_{-0.04}$  & -8.23 $\pm$ 0.35 & 2.19\\
UGC11648 & -23.53$^{+0.02}_{-0.03}$ & -3.24$^{+0.01}_{-0.01}$ & -113.35 $\pm$ 0.43 & 206.47 & -23.41$^{+0.02}_{-0.02}$ &  -3.26$^{+0.01}_{-0.01}$ &-315.56 $\pm$ 0.44 & 609.2\\
UGC11748 & -23.69$^{+0.01}_{-0.01}$ & -2.92$^{+0.01}_{-0.01}$ & -122.42 $\pm$ 0.46 & 465.26 & -23.63$^{+0.01}_{-0.01}$ & -2.92$^{+0.004}_{-0.004}$ & -244.21 $\pm$ 0.46 & 222.06\\
    \end{tabular}
    \caption{Parameter constraints, $\log{(E)}$ and  $-2\ln{\mathcal{L}}$ for each galaxy. The $\psi_{100}$ and $\psi_{210}$ cases have the same free parameters, $\log(m_{a})$ [eV/c$^{2}$] and $\log(\epsilon)$. The errors are reported with $1\sigma$.}
    \label{tab:chi2_results100_210}
    \end{ruledtabular}
\end{table*}

\begin{table*}
    \begin{ruledtabular}
    \begin{tabular}{c|ccccccc}
         
 \textbf{Galaxy} &$\log{(m_{a})}$ & $\log{(\epsilon)}$ & \textbf{$\log{(\psi_{100}(0))}$} & \textbf{$\log{(\psi_{210}(0))}$} &\textbf{$\log{(\psi_{320}(0))}$}  & $\log(E)$ & \textbf{$-2\ln{\mathcal{L}}$}\\ 
\hline
 ESO3020120 & -23.16$_{-0.15}^{+0.14}$  & -3.53$_{-0.05}^{+0.04}$& -0.01$_{-0.78}^{+0.01}$ & -3.41$_{-1.58}^{+3.37}$ & -0.94$_{-1.06}^{+0.46}$ & -7.28 $\pm$ 0.22 & 0.41\\ 
ESO3050090 & -22.74$_{-0.56}^{+0.17}$ & -3.64$_{-0.11}^{+0.51}$ & -0.09$_{-0.79}^{+0.09}$ & -1.59$_{-0.41}^{+1.57}$ & -0.54$_{-1.46}^{+0.48}$ & -6.74 $\pm$ 0.20 & 0.55\\ 
ESO4880490 & -23.01$_{-0.53}^{+0.07}$ & -3.46$_{-0.08}^{+1.02}$ & 0.07$_{-1.54}^{+0.10}$ & -3.38$_{-1.61}^{+3.20}$ & -0.53$_{-4.46}^{+0.49}$ &-9.82 $\pm$  0.24 &1.58\\
UGC11557 & -23.29$_{-2.35}^{+0.25}$ & -3.34$_{-0.32}^{+1.34}$ & -0.23$_{-1.78}^{+0.53}$ & -1.65$_{-2.35}^{+1.95}$ & 0.78$_{-3.22}^{+1.05}$ & -7.69 $\pm$0.11 & 0.35\\
UGC11616 & -23.27$_{-0.05}^{+0.05}$ & -3.25$_{-0.02}^{+0.25}$ & -0.01$_{-0.51}^{+0.01}$ & -1.73$_{-0.27}^{+1.19}$ & -0.95$_{-1.05}^{+0.21}$ & -18.80 $\pm$ 0.27 & 15.47\\
UGC4115 & -22.15$_{-1.75}^{+0.28}$ & -3.39$_{-0.52}^{+0.39}$ & -0.78$_{-0.60}^{+0.95}$ & -3.95$_{-4.10}^{+2.05}$ & -1.82$_{-1.98}^{+4.18}$ & -7.13 $\pm$ 0.11 & 0.01\\
ESO0140040 & -23.98$_{-0.09}^{+0.04}$ & -2.91$_{-0.02}^{+0.44}$ & -0.03$_{-0.83}^{+0.03}$ & -2.85$_{-0.15}^{+2.26}$ & -0.83$_{-2.17}^{+0.23}$ & -16.73 $\pm$ 0.02 & 10.47\\
ESO0840411 & -23.20$_{-2.80}^{+0.19}$ & -3.64$_{-0.13}^{+2.64}$ & 0.03$_{-3.21}^{+0.14}$ & -3.65$_{-2.35}^{+3.81}$ & -0.26$_{-5.74}^{+0.43}$ & -7.52 $\pm$ 0.11 & 0.21\\
ESO1200211 & -22.09$_{-1.42}^{+0.24}$ & -3.92$_{-0.27}^{+1.84}$ & -0.15$_{-3.78}^{+0.45}$ & -2.96$_{-2.04}^{+3.25}$ & -1.28$_{-3.72}^{+1.54}$ & -8.51 $\pm$ 0.16 & 1.21\\
ESO1870510 & -22.38$_{-0.44}^{+0.13}$ & -3.86$_{-0.09}^{+0.83}$ & 0.12$_{-1.40}^{+0.18}$ & -2.98$_{-1.02}^{+3.27}$ & -0.34$_{-3.66}^{+0.53}$ & -8.75$\pm$0.17 & 0.48\\
ESO2060140 & -23.26$_{-0.08}^{+0.07}$ & -3.29$_{-0.02}^{+0.36}$ & -0.001$_{-0.72}^{+0.001}$ & -3.66$_{-2.34}^{+3.19}$ & -1.02$_{-4.98}^{+0.38}$ & -22.04 $\pm$ 0.15 & 17.53\\
ESO4250180 & -23.69$_{-1.01}^{+0.28}$ & -3.18$_{-0.13}^{+1.18}$ & -0.12$_{-4.59}^{+0.12}$ & -4.91$_{-1.09}^{+4.89}$ & -0.90$_{-5.09}^{+0.69}$ & -8.90$\pm$0.02 & 0.95\\
F730-V1 & -23.38$_{-0.09}^{+0.10}$ & -3.22$_{-0.02}^{+0.37}$ & -0.006$_{-0.70}^{+0.01}$ & 1.88$_{-3.12}^{+1.34}$ & -0.96$_{-3.98}^{+0.27}$ & -22.56 $\pm$  0.21 & 18.82\\
UGC11454 & -23.46$_{-0.04}^{+0.04}$ & -3.20$_{-0.02}^{+0.32}$ & -0.01$_{-0.63}^{+0.01}$ & -1.69$_{-3.31}^{+1.00}$ & -0.83$_{-1.24}^{+0.18}$ & -32.10 $\pm$  0.23 & 35.19\\
UGC11583 & -22.02$_{-1.48}^{+0.18}$ & -3.66$_{-0.26}^{+1.65}$ & -0.55$_{-4.39}^{+0.55}$ & -0.74$_{-4.26}^{+0.73}$ & -2.26$_{-2.74}^{+2.21}$ & -7.93 $\pm$ 0.16 & 0.34\\
UGC11648 & -23.46$_{-0.04}^{+0.03}$ & -3.26$_{-0.02}^{+0.27}$ & -2.02$_{-0.54}^{+0.02}$ & -1.07$_{-3.93}^{+0.14}$ & -0.77$_{-1.06}^{+0.09}$ & -103.50 $\pm$  0.23 & 176.88\\
UGC11748 & -23.71$_{-0.01}^{+0.19}$ & -3.00$_{-0.07}^{+0.42}$ & 0.19$_{-3.96}^{+0.11}$ & -5.48$_{-0.48}^{+5.39}$ & -1.33$_{-4.66}^{+0.91}$ & -124.61 $\pm$ 74.57 & 224.84\\
 
    \end{tabular}
    \end{ruledtabular}
    \caption{Parameter constraints, $\log{(E)}$ and $-2\ln{\mathcal{L}}$ for the multi-state case of each galaxy. The free parameters $\log{m_{a}}$ [eV/c$^{2}$], $\log{\epsilon}$, $\log{\psi_{100}(0)}$, $\log{\psi_{210}(0)}$ and $\log{\psi_{320}(0)}$. The errors are reported at $1\sigma$.}
    \label{tab:nested_results_multistate}
\end{table*}

\ \\
\begin{turnpage}
\begin{table*}
\begin{ruledtabular}
\begin{tabular}{c|ccc|cccc|cccc}
       \multirow{2}{7em}{\textbf{Galaxy}}  & \multicolumn{3}{c|}{\textbf{$\psi_{100}$}}&\multicolumn{4}{c|}{\textbf{$\psi_{210}$}} & \multicolumn{4}{c}{\textbf{Multi-state}}\\
         
  & AIC & BIC & \textbf{$-2\ln{\mathcal{L}}$} & AIC & BIC & $\log(B_{12})$ & \textbf{$-2\ln{\mathcal{L}}$}& AIC & BIC & $\log(B_{12})$ & \textbf{$-2\ln{\mathcal{L}}$}\\ 
\hline
 ESO3020120 &  6.46 & 5.76   & 0.96 &  14.11 & 13.40  & -3.94 & 8.61 & 22.41 & 12.40 & 0.65 &0.41\\ 
ESO3050090 &  5.91 & 6.53 & 0.98 &  16.04 & 16.66 & -6.39 & 11.11 & 16.64 & 14.50 & 0.31 & 0.55\\ 
ESO4880490 &  8.30 & 7.59 & 2.80 & 30.20 & 29.50 & -11.98 & 24.70 & 23.75 & 13.74 & -0.93 &1.58\\
UGC11557 &  6.23 &5.12 & 0.52 & 13.33 & 12.22 & -4.95 & 7.61 & 25.33 & 11.84 & -1.29 & 0.35\\
UGC11616 &  28.64 & 28.57   & 23.44 & 111.17 & 111.1 & -56.26 & 105.97 & 35.21 & 29.46 & -12.40 & 15.47\\
UGC4115 &  5.24 & 5.17  & 0.04 & 9.97 & 9.90 & -4.22 & 4.77 & 18.58 & 12.84 & -1.38 & 0.01\\
ESO0140040 &  19.76 & 17.52   & 13.36 & 63.67 & 61.43 & -22.45 & 57.27 & 50.57 & 20.97  & -1.11 &  10.47\\
ESO0840411 &  6.45 & 4.84  & 0.45 & 18.21 & 16.6 & -6.99 & 12.21 & 30.21 & 11.19 & -1.04&  0.21\\
ESO1200211 &  6.65 & 6.84  & 1.56 & 9.6 & 9.78 & -1.01 & 4.51 & 18.71 & 14.41  & -0.94&  1.21\\
ESO1870510 &  6.32 &5.61  & 0.82 & 13.66 & 12.95 & -3.90 & 8.16 & 22.48 & 12.47  & -0.99 &  0.48\\
ESO2060140 &  28.36 & 28.77  & 23.36 & 94.96 & 95.37 & -33.18 & 89.96 & 34.20 & 31.07 &-0.73 &  17.53\\
ESO4250180 &  7.65 & 5.41  & 1.25 & 36.66 & 34.42 & -2.70 & 30.26 & 40.95 & 11.35  & -2.02 &  0.95\\
F730-V1 &  31.76 & 29.52  & 25.36 & 98.14 & 95.90 & -33.18 & 91.74 & 59.14 & 29.53  & -0.77 &  18.82\\
UGC11454 &  55.31 & 54.94  & 49.97 & 241.90 & 241.54 &  -92.83& 236.57 & 55.19 & 47.61  & 3.35 &  35.19\\
UGC11583 &  6.21 & 5.10 & 0.50 & 7.91 & 6.79 & -1.66 & 2.19 & 25.34 & 11.85 & -1.36 &  0.34\\
UGC11648 &  211.1 & 212.65 & 206.47 &  614.55 & 616.2 &-202.21 & 609.92 & 241.01 & 242.71 &9.85&  176.88\\
UGC11748 &  226.86 & 227.84  & 222.06 & 470.06 & 471.04 & -121.79 & 465.26 & 199.08 & 198.53 & 18.92 & 224.84\\
    \end{tabular}
    \caption{Results for model comparison between the models $\psi_{100}$, $\psi_{210}$ and multi-state for each galaxy.}
    \label{tab:model_comparison}
    \end{ruledtabular}

\end{table*}
\end{turnpage}


%%---------------------------------------------------------------------------------
\section{Conclusions and discussion} \label{Sec:Conclusions_and_discussion}
%%---------------------------------------------------------------------------------

This work analyzed the viability of $\ell$-boson stars as a dark matter component in the rotational curves, using bayesian statistics tools such as the NS along with the Bayes factor, together with the information criteria to know which case the data favours, being the multi-state case.
Seventeen LBSG were analyzed taking into account three $\ell$-boson stars cases, ground state, which was taken as the base model to calculate the Bayes factor, a single excited state and multi-states. In summary all galaxies prefer to be made up of multi-states as FIG \ref{fig:BayesFactor} suggests.
The plot's contours in FIG \ref{fig:RC_graphs} confirmed what the Bayes factor told us about the data, by obtaining a good parameter inference with the multi-state case. 
Although, there are some galaxies like UGC11648 and UGC11748, that are barely fitted but clearly they don't follow the model behavior. Additionally, it is noticed that for most galaxies the main contribution to the total rotational curve (dark blue line) is by the ground state ($\psi_{100}$, blue line) and the second excited state ($\psi_{210}$, orange line), this could be due to the spherical system assumption, meaning that more studies in this direction need to be done by solving the multi-state axial system of equations presented in \cite{Guzm_n_2020}. 
Furthermore, it is important to mention that by adding more states and therefore, increasing the free parameters, the scalar field mass tends to become slightly bigger than the ground state, this could be seen in the first columns of the tables \ref{tab:chi2_results100_210} and \ref{tab:nested_results_multistate}. Although it doesn't satisfy the constrains by Lyman-$\alpha$ \cite{PhysRevLett.119.031302_Lymanalpha_constrain,2019Lyman_Alpha_constrain}.\\
Although, the NS convergence for the multi-state case is not clear compared to the independent results, specially for the ground state central amplitude ($\psi_{100}(0)$) presenting an upper limit boundary at $\log{(\psi_{100}(0))}=0$, that could be interpreted as a necessity to extend the prior for this parameter. It can always be found a way to reparameterized it.\\
One of the main results that we can observe in FIG \ref{fig:Contoursl0}, FIG \ref{fig:Contoursl1} and FIG \ref{fig:Contoursl012}, is that the correlation between the parameters $m_a$ and $\epsilon$ seemed to be broken for the excited state ($\psi_{210}$) and the multi-state case. However, by looking at the posteriors in the triangle plots (FIG \ref{fig:Triangle_ESO3020120} and those in the \href{https://github.com/atalianb/Triangle_plots_ell_boson_stars}{repository}\footnote{\url{https://github.com/atalianb/Triangle_plots_ell_boson_stars}}), we can see that this correlation between parameters remains, with $\epsilon$ and the ground state central amplitude, $\psi_{100}(0)$.\\
Those correlations between parameters for dark matter density profiles for rotational curves have been studied by \cite{2017PhRvD..96d3005U}, where based on the mass discrepancy acceleration relation (MDAR) and that any DM halo will have a maximum acceleration they could break the correlation by a reparametrization, having just one free parameter. 
Another approach that could take place in the study to break the correlation between parameters is a reparametrization with the number of particles in each state by fixing the total number of particles and adding the ratios of the number of particles in different states with respect to the ground state \cite{2010PhRvD..82l3535U} to the boundary conditions of the shooting method.
However, by trying this approach we observed that the posterior distribution functions for the ratios of the number of particles where almost flat. Therefore, it is necessary to perform more studies in this regard, taking into account the approach in \cite{2017PhRvD..96d3005U}.\\
Perhaps, the two approaches mentioned above and adding the self interaction term in the potential could lead to a complete new analysis and to obtain a way that constrains the number of states.\\
It has been mentioned in the literature by \cite{2018MiguelAspeitia} and references therein, that adding more states could lead to a better parameter estimation but also, it is more expensive computationally. Because of this last reason we truncate the expansion to the third term, however we do a less expensive analysis, a $\chi^{2}$ fit (see \ref{app:ell3}), adding a fourth term in the expansion, for ESO3050090 and UGC11616 rotational curves. We find a better fitting and an important  contribution to larger radius in the rotational curve extension by the latest state. We obtained $\chi^2 = 0.31$ and $\chi^2 = 10.79$ for ESO3050090 and UGC11616 respectively. Which compared to the maximum likelihood obtained in the multistate case (i.e. where the expansion is truncated to the third term) is smaller. Additionally, the value of the scalar field mass is slightly bigger (see Table \ref{tab:ell3_results_parameters}). One can notice that in this case, the states that contribute the most to the rotational curve are the ground state ($\psi_{100}$, light blue line) and the last excited state ($\psi_{430}$, pink line), as we can observe in the Figure \ref{fig:RC_l0123}. While a physical explanation for this is still to be found, an statistical one could be obtain by comparing the AIC and BIC.\\
By comparing the AIC and BIC, respectively, it is observed a small increment in their values when including the $\psi_{430}$ terms (see Table \ref{tab:ell3_AIC_BIC_Chi2}), this could indicate that for the galaxy's sub-sample the multistate case is favored. Although, a more detailed analysis is needed in that direction and will be addressed in an upcoming work altogether with an improvement in the code optimization.

%%---------------------------------------------------------------------------------
\begin{acknowledgments}
%%---------------------------------------------------------------------------------

A.N.B. acknowledge financial support from CONACyT doctoral fellowship and
thanks the LSSTC Data Science Fellowship Program, which is funded by LSSTC, NSF Cybertraining Grant $\#$1829740, the Brinson Foundation, and the Moore Foundation; her participation in the program has benefited this work. The authors are gratefully for the computing time granted by Cluster Teopanzolco at Instituto de ciencias Físicas and the Instituto de Ciencias Nucleares at the Universidad Nacional Autónoma de México.
J.A.V. acknowledges the support provided by FOSEC SEP-CONACYT Investigaci\'on B\'asica A1-S-21925,UNAM-DGAPA-PAPIIT IN117723. A.B.B. and J.A.V. acknowledges the support provided by FORDECYT-PRONACES-CONACYT/304001/2019.
\end{acknowledgments}

\section*{Data Availability Statement}
The data underlying this article are available at \cite{Data}. The datasets were derived from sources in the public domain: \url{http://astroweb.case.edu/ssm/data/RCsmooth.0701.dat}.

\appendix
\section{$\chi^{2}$ results with $\psi_{100}$,$\psi_{210}$,$\psi_{320}$ and $\psi_{430}$}\label{app:ell3}
In this section we show the results obtained with the $\chi^{2}$ truncating the expansion in the fourth term in the system of equation SP (\ref{eq:dif_Schrodinger}-\ref{eq:dif_Poisson}), meaning that we are adding the terms and equations corresponding to $\psi_{430}$, the free parameters are $m_a$, $\epsilon$, $\psi_{100}(0)$, $\psi_{210}(0)$, $\psi_{320}(0)$ and $\psi_{430}(0)$. Where we have chosen the galaxy ESO3050090 and UGC11616 due to the value obtained for the maximum likelihood in the multistate case, described previously and their radial extension. 
The Table \ref{tab:ell3_results_parameters} displays the parameter estimation obtained from the $\chi^{2}$ for the galaxies ESO3050090 and UGC11616, and the Table \ref{tab:ell3_AIC_BIC_Chi2} contains the values for the AIC, BIC and $\chi^{2}$.

%%%%%
%
\begin{table}[htbp]
    \begin{ruledtabular}
\begin{tabular}{c|cc}
\textrm{\textbf{Parameters}} &\multicolumn{2}{c}{\textbf{Galaxy}} \\
& ESO3050090 & UGC11616\\
\colrule
\textrm{\textbf{$\log(m_a)$}} & -22.66 & -23.24 \\
\textrm{\textbf{$\log(\epsilon)$}} & -3.71 & -3.40\\
 \textrm{\textbf{$\log(\psi_{100}(0))$}}  & -6.30$\times 10^{-3}$ & 0.27\\
\textrm{\textbf{$\log(\psi_{210}(0))$}} & -1.46 &-2.74 \\
\textrm{\textbf{$\log(\psi_{320}(0))$}} & -1.40 & -1.30 \\
\textrm{\textbf{$\log(\psi_{430}(0))$}} & -0.60 & -0.59
\end{tabular}
    \end{ruledtabular}
\caption{Parameter estimation for ESO3050090 and UGC11616 with the $\chi^{2}$ for the multistate case with three excited states. The free parameters $\log{(m_{a})}$ [eV/c$^{2}$], $\log({\epsilon})$, $\log{(\psi_{100}(0))}$, $\log{(\psi_{210}(0))}$, $\log{(\psi_{320}(0))}$ and $\log{(\psi_{430}(0))}$.}
    \label{tab:ell3_results_parameters}
%%%%%%
\end{table}
\begin{table}[htbp]
\begin{ruledtabular}
\begin{tabular}{c|c|c|c}
\textrm{\textbf{Galaxy}} & \textrm{\textbf{AIC}} & \textrm{\textbf{BIC}} & \textrm{\textbf{$\chi^{2}$}}\\
      \colrule
ESO3050090 & 21.64 & 16.95 & 0.31\\ 
UGC11616 & 36.79 & 26.18 & 10.79\\ 

 \end{tabular}
 \end{ruledtabular}
\caption{Results for ESO3050090 and UGC11616 for the multistate case with three excited states.AIC, BIC and $\chi^{2}$.}
    \label{tab:ell3_AIC_BIC_Chi2}
\end{table}
%%%%%
%
\begin{figure}
         \centering
         \subfloat{\includegraphics[width=0.49\textwidth]{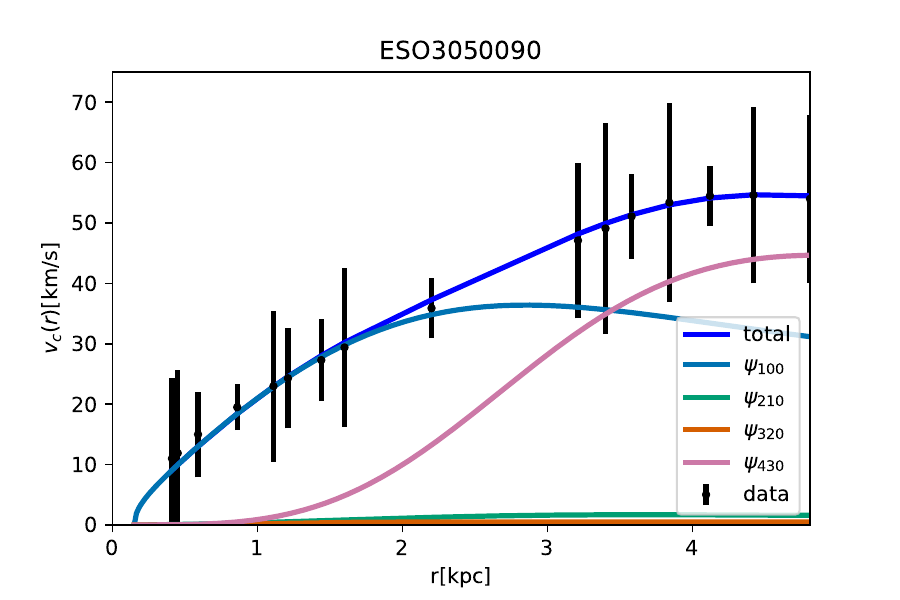}}
         \label{fig:ESO3050090_l0123}
     \hfill
         \centering
        \subfloat{ \includegraphics[width=0.49\textwidth]{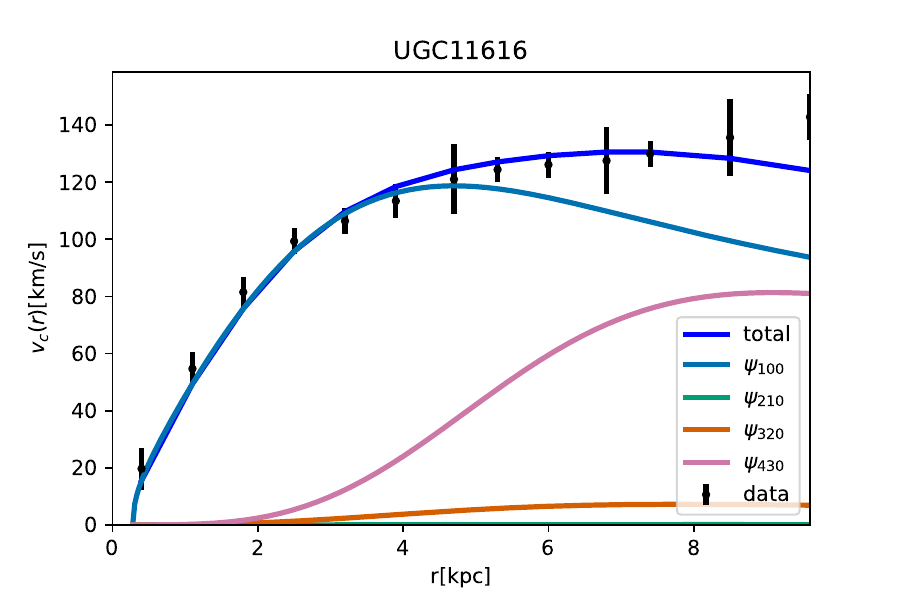}}
         \label{fig:UGC11616_l0123}
          
        \caption{Rotational curves for ESO3050090 and UGC11616 with the parameter estimation obtained from the $\chi^{2}$. The blue dark line indicates the resulting rotaitonal curve, the blue light line indicates the contribution of the state $\psi_{100}$, the green the contribution of the $\psi_{210}$, the orange the contribution of $\psi_{320}$ and the pink the contribution of $\psi_{430}$.}
        \label{fig:RC_l0123}
\end{figure}

%\subsection{\label{app:subsec}A subsection in an appendix}

%You can use a subsection or subsubsection in an appendix. Note the
%numbering: we are now in Appendix~\ref{app:subsec}.

%Note the equation numbers in this appendix, produced with the
%subequations environment:
%\begin{subequations}
%\begin{eqnarray}
%E&=&mc, \label{appa}
%\\
%E&=&mc^2, \label{appb}
%\\
%E&\agt& mc^3. \label{appc}
%\end{eqnarray}
%\end{subequations}
%They turn out to be Eqs.~(\ref{appa}), (\ref{appb}), and (\ref{appc}).

\nocite{*}

%\bibliographystyle{mnras}%\bibliographystyle{unsrt}%{abbrv}
%\bibliographystyle{ieeetr}
%\bibliography{apssamp}

\end{document}